\documentclass[10pt, final, journal, letterpaper, oneside, twocolumn,romanappendices]{IEEEtran}

\pdfoutput = 1
\usepackage{ifpdf}
\usepackage[english]{babel}
\usepackage{amsmath}
\usepackage{amssymb}
\usepackage{accents}
\usepackage{fixltx2e}
\usepackage{graphicx}
\usepackage{fullpage}
\usepackage{enumerate}
\usepackage{enumitem}
\usepackage{gensymb}
\usepackage{undertilde}
\usepackage{mathrsfs}
\usepackage{setspace}
\usepackage{cite}
\usepackage{units}
\usepackage[bottom]{footmisc}
\renewcommand{\footnoterule}{%
  \kern 3pt
  \hrule width 1in height .5pt
  \kern 3.5pt
}

\title{Security of Quantum Key Distribution}

\author{Horace~P.~Yuen}

\date{}

\begin{document}
\linespread{1}
\maketitle
\linespread{2}
\begin{abstract}
          
The security issues\footnote{\noindent This paper is to appear in IEEE Access} facing quantum key distribution ​(QKD) are explained, herein focusing on those issues that are cryptographic and information​​ theoretic ​in nature​ and not those based on ​physics​. T​he problem of ​security criteria is addressed. It is demonstrated that an attacker's success probabilities are the fundamental criteria of security that any theoretic security criterion must relate to in order to have operational significance. The errors committed in the prevalent interpretation of the trace distance criterion are analyzed. The ​security proof​s of QKD​​ ​protocols are discussed and assessed in regard to three main features: their validity, completeness, and adequacy of the achieved numerical security​ ​level. Problems are identified in all these features. It appears that the QKD security situation is quite different from the common perception that a QKD-generated key is nearly perfectly secure. Built into our discussion is a simple but complete quantitative description of the information theoretic security of classical key distribution that is also applicable to the quantum situation. In the appendices, we provide a brief outline of the history of some major QKD security proofs, a rather unfavorable comparison of current QKD proven security with that of conventional symmetric key ciphers, and a list of objections and answers concerning some major points of the paper.

\end{abstract}PACS \#: 03.67Dd

\section{Introduction}
Quantum key distribution (QKD) involves the generation of a shared secret key between two parties via quantum signal transmission [1], [2]. (Among other possible terms, we will often use the more appropriate ``generation" in lieu of ``distribution," ignoring their fine distinction in conventional cryptography [3].) QKD is widely perceived to have been proved secure in various protocols [1], [2], in contrast to the lack of security proofs for conventional methods of encryption for privacy ​or ​key distribution.​ Security proofs in ​QKD are​ highly technical​ and ​are ​​also ​multi-disciplinary in nature, as is the case with the subject area of quantum cryptography itself. Theoretical QKD involves in its description and treatment various areas in ​quantum physics, information theory, and cryptography ​at​ an abstract  and conceptual level. It is difficult for non-experts in QKD security to make sense of the literature; moreover, even experts are often not aware of certain basics in some of the relevant fields. 
\IEEEpubidadjcol Many who perform assessments on QKD security  ​follow the vague community consensus on QKD security being guaranteed by rigorous proofs. A common perception is that QKD gives ``perfect secrecy," as asserted for example in a useful recent monograph on conventional cryptography [3, p. 589]. It is interesting to \vspace*{\fill}
\thanks{This paper is to appear in IEEE Access}
note that QKD is commonly taught to physics students as being an important application of quantum optics because QKD is provably secure. To our knowledge, the provable security property is often taught as being self evident and is not questioned on any level (recent advances in quantum hacking may be an exception; however, such attacks are based on discrepancies between the model and real systems as opposed to the security of the model itself). The commonly cited reason for no-cloning or quantum entanglement is very far from sufficient. Even in the technical literature, a QKD-generated key is often regarded as perfect whenever it is used in an application. One main purpose of this paper is to correct such a misconception and to demonstrate how the imperfect generated key affects the security proofs themselves.

Security proofs by their nature are conceptual, logical and mathematical yet indispensable for guaranteeing security. ​Cryptographic security  cannot b​e guaranteed​ b​y ​experiment, if only because possible attack scenarios cannot be exhausted via experiment. However, security is a most serious issue in cryptography and must be thoroughly and carefully analyzed [4, two prefaces and ch. 1; see also quotes in Appendix I]. \textit{The burden of proof} is on one who makes the security claim, not on others to produce specific successful general attacks. 

In this paper, we will describe the actual security theory situation of QKD with just enough technical​ ​materials for accurate statements on the results. ​We will be able to describe some main security issues without going into the physics, and we can treat everything at a \textit{classical} probability level, to which a \textit{quantum} description invariably reduces. We will discuss in what ways these security issues have been handled inadequately. Some major work in the QKD security literature will be mentioned  and also discussed in Appendix I, which may help clarify the issues and illuminate the development that led to the current security situation. In Appendix II, we compare QKD to conventional cryptography and provide a preliminary assessment on the usefulness of QKD when conventional cryptography appears adequate. (Note that cryptography is a small and relatively minor subarea of computer security [4]. It is the latter that results in news headlines.) In Appendix III, some possible objections to certain points of this paper from the viewpoint of the current QKD literature are answered. Table 1 in Section VIII.B gives a summary comparison of various numerical values.

Generally, perfect security cannot be obtained in ​any ​key distribution scheme with security dependent on physical characteristics due to system imperfections mixed with the attacker's disturbance, which must be considered in the security model. This is especially the case with QKD, which involves small signal levels. We use the term ``QKD" in this paper to \textit{refer} to protocols with security depending on information-disturbance tradeoffs [1], [2], excluding those based on other principles such as the ``KCQ" approach in [5], which permits stronger signals and for which no general security proof has yet been claimed. In QKD, one can at best generate a key that is close to perfect in some sense.​ ​This immediately raises the issue of a security criterion, its operational significance and its quantitative level. Security is very much a quantitative issue. Quantitative security is quite hard to properly define and to rigorously evaluate; thus, there are few such results in the literature on conventional mathematics-based cryptography. It is at least as hard in physics-based cryptography, and there is yet no true valid quantification of QKD security under all possible attacks. 

That there are problems and gaps in QKD security proofs has been discussed since 2003 in [6, App. A], [7], [5, App. A and B], [8], culminating in the numerical adequacy issue in [9] in 2012, which provides the trace distance criterion level for a so-called ``near-perfect" key. This last numerical adequacy point is emphasized in [10], and a reply is given in [11], which in turn is replied to in [12]; no further exchange on this topic has resulted. The basic point of [11] is that a trace distance level of $10^{-20}$ is sufficient for security. There have since been several arXiv papers that elaborate upon the several QKD security issues that have yet to be resolved. This paper summarizes and supersedes those papers in a coherent framework for analyzing QKD security. This paper shows in what ways, even at a value of $10^{-20}$, which is ten orders of magnitude beyond what is currently achievable, such a trace distance level does not provide adequate security guarantees in a cryptosystem that involves the use of a vast number of keys at such a level; see Section VIII.

Before proceeding to a detailed treatment, we list the major problems of currently available QKD security proofs as follows:

\begin{enumerate}[label=(\roman*)]
\item The chosen security criterion, namely, a quantum trace distance $d$, has been misinterpreted. The operational security guarantee that it yields does not cover some important security concerns.
\item The numerical security level that has been obtained is far from adequate. The very strong level of guarantee asserted recently is derived from erroneous reasoning.
\item The known security proofs are not complete nor justified at various stages in a valid manner, especially in connection with the major necessary step of error correction, which has not been treated in a rigorous manner.
\item A trace distance level guarantee of a key $K$ limits the information-theoretic security level that can be obtained when $K$ is used for message authentication, taking away an otherwise available security parameter that allows arbitrarily high levels of message authentication security.
\end{enumerate}

There are many other serious issues facing QKD security proofs, many of which relating to physics and implementation. These issues will not be discussed in this paper, which concentrates on a careful exposition of the above four points.

Much of the technical content in this paper is conceptual analysis, especially on the use of probability in real-world applications. The applications are not essentially mathematical or physical in nature, which is partly why they are easy to miss and result in various confusions. Sections III.A, IV, V.C, VIII.B, and Appendix III contain relevant clarifications on the subtle meaning of probability in real-world applications. Note that knowledge of physics, classical or quantum, is \textit{not} required to understand the content of this paper. Furthermore, the relevant basic cryptography concepts will be explained when being introduced.

\section{Conventional and Quantum Cryptography}

We briefly review the different representations of conventional and quantum cryptography in regard to the cryptographic goals of privacy and key distribution [3]. 

\begin{figure}[!t]
\centering
\includegraphics[width=3in]{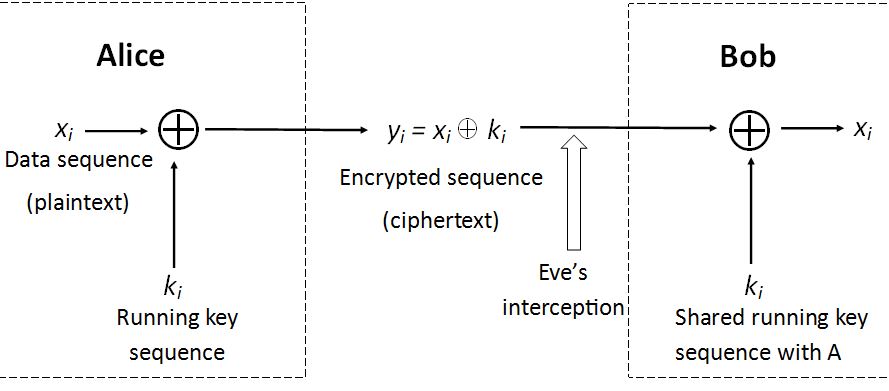}
\caption{Additive stream cipher in which the running key sequence $\{k_i\}$ may or may not be uniformly distributed.}
\end{figure}

​​In Fig. 1, the conventional stream cipher encryption of a data sequence $x = \{x_j\}$ for privacy ​is depicted. A user Alice transmits $y=\{y_i\}$, which is the xor of the data bits $x_i$ and running key bits $k_i$ for each $i$:

\begin{equation}
Y_i = X_i \oplus K_i 
\end{equation} We use uppercase to denote random variables and lower case to denote the values they take on. Thus, from (1), Alice transmits $y_i = x_i \oplus k_i$ for given $x_i$ and $k_i$. A prior shared key bit sequence $k = \{k_i\}$ unknown to the attacker Eve is known to Alice and the other user Bob, who can decrypt $x_i=y_i\oplus k_i$ from $y_i$ by knowing $k_i$. The attacker Eve would ​then ​learn nothing about $X$ ​from $Y$ without knowing something abo​ut $K$. ​​She knows nothing about $K$ when ​the $K_i$ are (statistically) independent bit​s​ with equal probability ​of being 0 or 1. ​In this​ so-called ​``​one-time p​ad" encryption​, her probability of obtaining the sequence $X$ correctly given that she knows $Y$ through interception is equal to her \textit{a priori} probability on $X$. For uniformly distributed $X$,

\begin{equation}
p(x|y)=2^{-n}
\end{equation} where $n = |X| = |Y| = |K|$ is the bit length of the sequences. (The vertical bar $|\cdot|$ is always used in this paper to denote the bit length of the sequence within it. Lower case $p(\cdot)$ are discrete probability distributions, and no continuous distributions will be used.) We may represent this by $K = U$, namely, the uniform random variable to Eve. Thus, in addition to $p(x_i|y_i)=1/2$, ​which is ​the \textit{a priori} probability of each $X_i$, there is also no correlation of any type between the $X_i$s that Eve can find. This is ``perfect secrecy." The security under discussion is the \textit{information-theoretic security} of the intrinsic uncertainty to Eve. (See Section III and Appendix II.) The correlation among bits in $K$ is a most important feature often missed in cryptography, especially in connection with the trace distance (statistical distance) criterion, which is presented in Section IV.A and also in Section IV.B in the context of distinguishability.

The goal of QKD is to generate a key $K$, which ideally is $K=U$, by transmitting bits $X'$ from Alice to Bob via quantum signals with no use of shared secret keys. In reality, a prior shared secret key is needed to start executing the QKD protocol, at least for message authentication against man-in-the-middle attacks.   

\begin{figure}[!t]
\centering
\includegraphics[width=3in]{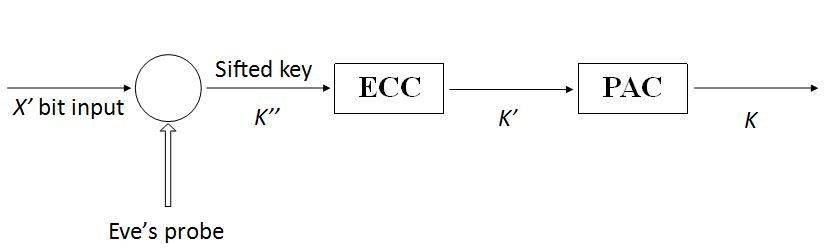}
\caption{Schematic representation of a QKD system incorporating error correction and privacy amplification with generated key $K$.}
\end{figure}

The QKD key generation process is depicted in Fig. 2.  For definiteness, the​ original​ BB84 protocol [1] is schematically described in the following, which contains the key idea of QKD. A sequence of quantum​ ​optical​ ​signals is modulated by ​the data $X'$ and sent from Alice to Bob, with each bit $X_i'$ modulating a separate quantum signal. (We use $X'$ to distinguish these key generation data from the data $X$ on which the QKD key $K$ is used to encrypt, as in Fig. 1.) In BB84, each quantum signal is a single photon in a so-called ``qubit​"​, a two-dimensional quantum state space. Eve could intercept and set her probe on the qubits during signal transmission. Bob measures on one of the two BB84 bases randomly upon receiving each qubit signal and obtains a bit value of 0 or 1. After the entire sequence is measured, Bob publicly announces which basis he measured on each $i$, and the ones ``mismatched" to Alice's ​transmitted signal are discarded. Then, a portion of the remaining matched ones is used to check the frequency of bit error, which is called the \textit{quantum bit error rate} (QBER). The other portion is called the \textit{sifted key} $K''$ from which the final key $K$ is to be generated.

For our purposes, there is ​\textit{no ​n​eed} to understand ​the​ exact​ physics and underlying rationale of the above proced​ure​. The only thing that matters for the purposes of this paper is the representation of Eve's knowledge on the generated $K$ via her final classical observed value $Y$ by the joint distribution $p(k,y)$. Eve's probe inevitably disturbs the quantum signals if she​ ​learns anything from the probe, a characteristic quantum effect of the information-disturbance tradeoff, which has no classical analog. It is usually assumed that the users ​would regard​ all disturbance as indicated by the QBER level to be from Eve's interception. They would estimate how much ``information" Eve can learn about $X'$​ with such a disturbance​. (The vague word ``information" would be specified precisely in context.) ​The users have to correct the errors in $K''$ to obtain a useful key​ $K$. Such errors would always be present from system imperfections due to the low signal level. Error correction is typically accomplished by an ordinary error correcting code (ECC) on $K''$, as indicated in Fig. 2. The users then transfer the estimate of Eve's information on $K''$ to the error-free $K'$. If ​Eve's information ​is below a certain threshold level, the users may employ ​``​privacy amplification​"​ [1] ​on $K'$ ​to eliminate it. The privacy amplification code (PAC) usually involves linear hashing compression [​3​] on $K'$ to a final generated key $K$ with bit length $|K|$, which is a small fraction of $|K'|$.

What is the security desired and claimed for the QKD-generated key $K$? Because this is a physical and in particular a quantum cryptosystem, there are many mutually exclusive different observed values $Y$ that Eve could obtain from her choice of quantum probe and based on the quantum measurement on the probe. She could estimate various properties of $K$ from the $Y$ that she obtains, each with a certain probability of success. She also would gather side information relevant to improving her estimates before she measures her probe and makes estimates. Such side information would include the BB84 bases open announcement and the specific PAC employed in the QKD​ ​round. It may also involve the specific ECC used. The users' goal is to make Eve's probability of success in obtaining any characteristic of $K$ close to the level whereby $K$ is perfect, i.e., when $K=U$.

An especially significant attack on privacy encryption is the \textit{known-plaintext attack} (KPA), which is the main vulnerability of conventional mathematics-based encryption. The \textit{ciphertext-only attack}, for which $X=U$ to Eve, that QKD security analyses focus upon is usually not considered as a serious risk for symmetric-key ciphers [3], [4], [8], which can be further substantiated in an information-theoretic manner. A brief summary is given in Appendix II, which compares QKD with conventional cryptography. A KPA runs as follows. 

Eve may know a portion of the data $X$ that is encrypted with $K$ and hence knows part of $K$ because $Y$ is open. She may then learn something about the remainder of $K$ through correlations among the bits in an imperfect $K$ and hence something about the unknown portion of $X$. It is such a KPA when a QKD key is being used that must be protected against. This implies that correlations between bits in $K$ must be addressed in QKD security.

Thus, \textit{p​erfect s​ecrecy} against all attack possibilities would require that $K$ is uniformly distributed to Eve for any $Y$ that she may possibly obtain that is allowed by the laws of quantum physics together with all​ her side information. Thus far, a security criterion is chosen to measure the difference between an imperfect key and a perfect key with a quantitative security level. The term ​\textit{u​nconditional ​security} was coined [13] and widely used  ​to include the following two conditions:\\
\text{ } \textit{Unconditional Security:}
\begin{enumerate}[label=(\roman*)]
\item complete generality on ​possible ​attacks;
\item quantitative security level can be made as close to perfect​ as​ desired.
\end{enumerate}
The data bit length $|X'|$ or $|K|$ in a QKD round is often taken to be a \textit{security parameter} $s$, namely, the quantitative security level improves with increasing $s$ and becomes perfect as $s\to\infty$ [13]. ​

Such​ ​unconditional security for QKD is often claimed because it distinguishes QKD from conventional cryptography; moreover, it is the only advantage of QKD (See Appendix II). Security against only some attacks means that the QKD approach may not lead to good security in the future when other attacks become practical. The latter is often taken to be the situation with conventional cryptography. Unconditional security in QKD remains asserted on occasion, both theoretically for a given protocol and physical models and experimentally as potential or even actualized possibility. However,  \textit{no} security parameter has ever been found for any QKD protocol at any key generation rate, and certainly, $|X'|$ and $|K|$ are not such parameters, as will be shown in Section III.B. 

To explain the process and requirements of a security proof, we proceed to quantitatively describe the information-theoretic security of a cryptosystem.​

\section{Operational Security Level of a Secret Bit String}

It is sometimes asserted that a cryptographic security criterion is a matter of ``definition" and ``interpretation," although this is highly misleading and can be considered incorrect. There are definite specific characteristics in a cryptographic goal on which users want to protect against successful attacks. A security criterion would be inadequate for a security task it is supposed to serve if it does not lead to a guarantee at an adequate quantitative level. The range of adequate levels may depend on the specific application; however, one cannot sensibly ``define" a protocol to be secure if its security criterion and level do not cover such possible attacks. ``Interpretation" of a mathematical statement, on security or any other matter, may be correct or incorrect when it is applied to a real-world situation. Moreover, there have been many erroneous interpretations in QKD security analysis. These errors are mainly conceptual errors, not mathematical nor mainly mathematical mistakes. These errors often involve reading ordinary meanings into a word that in context is a technical term that carries only a precise technical meaning. We will see many examples of these errors in Sections III to VIII of this paper.

The operationally or empirically meaningful security criteria on the secrecy of any shared key string $K$ for privacy or key distribution, whether it is generated by QKD or any method, ​are​ the attacker Eve's probabilities or rates of success in correctly obtaining ​various​ parts or characteristics of $K$, including $K$ itself in its entirety. This is the case even in complexity-based security, as we will see.  ​

\subsection{Why Probability Criteria are Needed}

The​ quantitative information-theoretic security of a key $K$ is often described by a single-number security criterion, such as Eve's Shannon mutual information $I(K;Y)$ on $K$ [14] through her observation $Y$, which she may obtain by intercepting the signal transmission. This $I(K;Y)$ is defined [14] from the joint distribution $p(k,y) = p(y|k)p(k)$, where $p(k)$ is the distribution of the generated $K$ obtained by the users under a given quantum probe from Eve. The conditional probability $p(y|k)$ is specified through the cryptosystem representation and Eve's measurement result, from which she derives her estimate of a characteristic $C(k)$, denoted by $\hat{C}(k)$, which is a function of $y$. For example, from $y$, she could estimate $C(k) = k$ as $\hat{K}(y)$. Because she takes $\hat{K}$ to be $K$, we can simplify our notation by simply writing $K$ itself instead of $\hat{K}$, i.e., $K$ is being observed from Eve's viewpoint. Thus, $p(y|k)$ gives the conditional probability that the  $y$ observed by Eve given $k$ is the actual key generated. Eve can now derive $p(k|y)$ for all possible $k$ through $p(y|k)$ and Bayes' rule.

The criterion $I(K;Y)$ gives the number of Shannon bits $I(K;Y)$ ​concerning​ $K$ known to Eve​ because, in this case, one can write

\begin{equation}
 I(K;Y) = |K| - H(K|Y)
\end{equation}
where $H(K|Y)$ is the conditional entropy of $K$ given $Y$ [14]. Note that, as we just indicated, Eve has a \textit{full distribution} on her knowledge of $K$ given any observation $y$ with the conditional distribution $p(k|y)$. We may assume that all ​the ​side information ​that ​Eve may possess has been considered in her final $p(k|y)$. We order the $N=2^{|K|}$ possible values of $K$ entering $p(k|y)$ and suppress its $y$ dependence so that, in various abbreviated notations, Eve has $p(k|y) \equiv \{p_i\} \equiv P$ with

\begin{equation}
p_1 \geq p_2 \geq ..... \geq p_N
\end{equation}
This probability profile is the complete ``information" Eve has on $K$ given her observation $y$. Any single number criterion, such as $I(K;Y)$, \textit{merely} expresses a constraint on $P$. When $I(K;Y) = 0$, ​we have $P=U$, and Eve knows nothing about $X$. For a given level $I(K;Y) = \epsilon > 0$, what does this imply ​for the​ ​security of $K$?​

This question arises for any criterion that is used as a theoretic quantity. A single-number information-theoretic measure on ​the ​uncertainty or ``information" is a theoretical quantity whose operational or empirical meaning needs to be independently explained [15, preface]. In the context of ordinary communications, the two theoretic quantities entropy and mutual information are related to the empirical data rate and error rate through the Shannon Source and Channel Coding Theorems. What would be the operational meaning of these quantities in the context of cryptography? One ​\textit{cannot} simply assume the word ``information" for a technical concept would carry its ordinary meaning in any application, especiall​y ​not quantitatively. Shannon himself emphasized such a danger early on [16]. In Shannon's cryptography paper [17], he used such information measures; however, except for the ideal case of $I(X;Y) = 0$ for a one-time pad, he did not explain their operational cryptographic significance.

In cryptography, one is concerned that Eve should not be able to correctly estimate various quantities associated with a key $K$ from her observation and side informat​io​n. ​​Such s​uccess​ i​s ​genera​lly​ ​obtained only probabilist​ically​. Therefore, this operational security requirement translates ​to Eve \textit{not} being able to estimate such quantities well, i.e., not with appreciable probability. In the case of perfect secrecy, Eve's $P$ above ​equals​ the uniform $U$. Therefore, in the general imperfect case​, her estimate probabilities as derived from $P$ should be close to that derived from $U$. In particular, the exact level needs to be quantitatively compared to that from $U$ and its numerical adequacy to ensure security for a given application. 

Thus, given a security criterion level that sets a constraint on $P$ above, we would need to ascertain what success probabilities Eve may possibly obtain. Specifically, we would compare the $p_1$ from any $p(k|y)$ not ruled out by the security criterion constraint to the $K=U$ level:
\begin{equation}
p_1(K) \text{ } \text{ vs } \text{ } U(k)=2^{-n}
\end{equation} 
We use the notation $p_1(K)$ to  explicitly demonstrate that the $p_1$ level refers to the $K$ with a distribution $p(k|y)$. Clearly, $p_1(K)$, Eve's maximum probability of correctly obtaining $K$, needs to be sufficiently small for any meaningful claim to security even if it may be far larger than the $2^{-n}$ level.​ We would address such a general security probability guarantee based on security measures in Sections IV and V. Their connections to the numerical security levels of concrete protocols are discussed in Section VIII.

When $K$ is used to encrypt data $X$​, part of $​X$​ may be known to Eve in a KPA, as discusse​d​​ in​​ ​Sec​t​ion​ II. Let $K_1$ be the subsequence of $K$ known to Eve, say, when $K$ is used as a one-time pad, and let $K_2^*$ be a subsequence of $K_2$, $K_2$ being the remainder of $K$ excluding $K_1$. Then, Eve's optimal success probabilities from such an attack should be compared to the perfect security level when $K=U$,

\begin{equation}
p_1(K_2^*|K_1=k_1) \text{ } \text{ versus } \text{ } p_1(U_2^*|U_1=u_1)=2^{-|U_2^*|}
\end{equation}
where $U_1$, $U_2$, and $U_2^*$ are the sequences obtained from $U$ with the same bit positions as $K_1$, $K_2$, and $K_2^*$, respectively. In general, Eve may possess only statistical information on $X$ without knowing part of it exactly. We will not address this more complicated situation in this paper.

It is important to note that we may write 
\begin{enumerate}[label=(OG)]
\item operational guarantee for an event = \\rule out its possible occurrence with a high probability
\end{enumerate}
An average number of occurrences (sample mean) is \textit{not} an operational guarantee because the number of occurrence is itself a random quantity from a finite number of trials, each of which has the same probability distribution. This does not mean that an average cannot be used as a measure of security. It means  that an average is a less accurate measure compared to a probability statement on individual occurrences or on a multiple-use sample mean. If the variance is known in addition to the average (mean), the probability statement on the sample mean can be made, and operational guarantees can be restored. If only the average is known, the Markov inequality can be used to provide an accurate individual probability statement, as shown in Section V.C.

Complexity-based security is operationally equivalent to the following success probability characterization. Let $M$ be the total number of possibilities that Eve can attempt computationally among the $N=2^n$ possible $K$ values to determine if a particular value is the correct value, as in opening a safe. It is easy to show that, ​with​ a uniform chance of success for each trial, her overall probability of success is, for a uniform probability distribution on the $N$ possible cases,

\begin{equation}
p_1(K) = p(k) = M/N  
\end{equation}
Eq. (7) can be readily generalized when Eve's success probability for each trial is not uniform but known [5]. Indeed, it can be observed that there is no difference between complexity-based security and information-theoretic (probabilistic) security if Eve is given a sufficient number of attempts to determine the correctness of a given $k$, as in the case of cracking a safe. She would need at ​m​ost $N$ trials and on average $N/2$ trials. The problem of complexity-based security is that it is very hard to prove a lower bound on the number of trials Eve needs for a given problem, and no such proof exists for any common problem. It will be observed that it is also very hard to prove QKD security, and no such proof yet exists as well​.​

\subsection{The Mutual Information Criterion}

In the literature on classical noise-based key generation within information-theoretic security [18], [19], [20]​,​ both before and after the emergence of QKD in 1984 [21], the security criterion used is the mutual information $I(K; Y)$ between the generated key $K$ and Eve's observation $Y$.  ​No relation of this information-theoretic quantity to ​Eve's ​opera​tional ​success ​probability was given until [5]​​. Th​e issue ​will be discussed further in the next section in connection with the statistical distance criterion. Here, we would like to remove a major misconception about security proofs that use the mutual information criterion, first discussed in Appendix A of ref [5] and which directly carries over to the QKD case for the $d$ criterion as well [5], [22].

Apart from the problem of bounding $p_1(K)$ from $I(K;Y)$, the latter we will abbreviate as $I_E$, ​the asymptotic security proofs that show, with $|K| = n$,

\begin{equation}
I_E \to 0 \text{ } \text{ } \text{ as } \text{ } \text{ } n \to \infty  
\end{equation}
were erroneously supposed as proofs that $K$ is asymptotically perfect. That is confusing the meaning of a limit because $\infty$ is  not a number. What occurs here is that the convergence rate of $I_E$ to $0$ determines the asymptotic security level as follows. From Lemma 2 in [5], for any $l<n$, it is possible that

\begin{equation}
{I_E}/{n} \sim\ 2^{-l} \text{ }\text{ } \text{ with } \text{ } \text{ } p_1(K) \sim\ 2^{-l}
\end{equation}
Eq. (9) states that, under the constraint of a given level of $I(K;Y)$, there are possible $p(k)$ with $p_1$ at the same quantitative level as ${I_E}/{n}$. Thus, a very insecure $K$, compared to $U$, can satisfy (8), even when $I_E$ converges exponentially in $n$:

\begin{equation}
I_E = 2^{-(\lambda n -\log{n})} 
\end{equation}
for a constant $\lambda$. It is possible given Eq. (10) that $p_1(K) \sim\ 2^{-\lambda n} \text{ for } \lambda \ll 1$, as compared to $2^{-n}$ for a uniform key. Apart from condition (i) of unconditional security in Section II, such an asymptotic proof of (8) does not imply condition (ii) for unconditional security. Indeed, it does not even imply $K$ is in any sense near perfect, as the above case (10) shows. We may mention that the current quantum criterion $d$ suffers from the same exponential problem as will be discussed in the following section. Although $d$ can be directly used in a finite $n$ protocol,​ ​this exponential problem is why relatively large and insecure levels of $d$ are obtained in a real protocol with sizable key rates.

The​ quantum generalization of $I_E$ is called ``accessible information", which is the maximum mutual information $I(K;Y)$ that Eve can obtain from any quantum measurement on her probe.​ ​Such an additional issue of measurement optimization is characteristic of quantum detection [23], [24]. This issue plays no role in our context after we take $I_E$ to be Eve's accessible information so that the quantum security situation reduces to a classical one under such $I_E$. The early proofs of QKD security up until 2004 are based on the use of such an accessible information criterion as well as the current proof on the so-called measurement-device-independent approach [25]. Thus, the proofs all suffer directly from the problems explained above and remain in error even after the proof is converted to one with the trace distance criterion. A particularly influential early security proof is given in [26], which is the basis of the  heuristic generalizations used to include various system imperfections in [27]. The side information that Eve has from the open announcement of ECC and PAC of Fig. 2 are considered in the proofs of [13], [26], [27]. In Section VII, we will discuss how they considered in more recent proofs and what problems are yet to be resolved.​

\section{The Trace Distance Security Criterion $d$ and its Security Meaning}

The mutual information criterion​ does not directly guarantee security against known-plaintext attacks (KPA). We require bounds on the conditional probability (6) when part of $K$, namely, $K_1$, is known to Eve so that correlation between the bits in $K$ will not leak much information about $K_2$, namely, the remainder of $K$. In QKD, this KPA problem is considered as one of ``universal composition" [28], [29], [30] in which the security when $K$ is being used in an application is taken to be a ``composition" security issue. Although KPA security is crucial and is the usual concern of privacy in conventional ciphers, as noted above, it was not addressed or discussed in the QKD literature until ref [28] twenty years after ref [21] appeared. This topic was addressed in [28] as a composition security issue, with the conclusion that security is ensured when the accessible information goes to $0$ exponentially in $|K|$. That is directly contradicted by (10) above even simply for ciphertext-only attack security.

Then, in ref [31], it was noted in a specific counter-example that a single-bit KPA leak is possible under the accessible information criterion due to the phenomenon of ``quantum information locking", and the trace distance criterion $d$ was proposed as an improved criterion (the $d$ criterion was also discussed in [28]) with the claim that, under $d\leq\epsilon$, the probability that $K$ is not perfect is at most $\epsilon$. Specifically, $d$ is claimed to be the maximum probability that the generated $K$ is not perfect, such probability being called the \textit{failure probability.} Apparently, the accessible information criterion is much worse. Under such a criterion, knowing $\log|K|$ number of bits in $K$ may fully reveal the remainder of $K$ [32].

Such a maximum failure probability interpretation of $d$, as originally given in [29], [30] and continuously maintained in many subsequent papers and in the general review [2], is \textit{incorrect}; however, it has been maintained publicly in the QKD community to date, despite its flaw having been revealed and explained in early 2009 [33], [34], in [5] and [8], and in several arXiv papers until the Fall of 2014 in [22]. Only in ref [35] i​s such an interpretation vaguely combined with a correct security consequence of $d$ (Eq. (14) below) but with no acknowledgment of previous errors. Part of the reason is likely that the ``indistinguishability advantage" interpretation of $d$ is employed instead for validation of this incorrect interpretation, which serves to justify QKD security that cannot be otherwise obtained. In Sections IV and V, we will treat this issue in detail to identify the major security issues involved and those that have not been resolved with the $d$ criterion. 

There are \textit{two} different derivations of the failure probability interpretation of $d$ in the QKD literature, which we will treat in Sections IV.A and IV.B. This incorrect failure probability interpretation of the QKD security criterion $d$ is prevalent, and the ``distinguishability advantage" derivation in Section IV.B remains widely quoted and discussed as validation of the interpretation. The issue is of major importance because the criterion issue and its ramifications lie at the foundation of  information theoretically secure key generation, both classically and in the quantum case. Thus, the full treatment of this issue in this section is very much warranted.

We proceed by first explaining how a security criterion functions in a physical cryptosystem where signal transmission can be intercepted. As described in Section III.A, based on her attack and the physical system representation, Eve could derive a conditional probability distribution $p(k|y)$ on the various possible values of $K$ given her observation $Y$. She also has side information from the execution of the protocol, namely, the public announcements in the QKD case, which we can label as $z$. We use the following notational abbreviations by suppressing the $y$ and $z$ dependences

\begin{equation}
p(k|y,z) \to p(k|y) \to p(k) = \{p_i\}
\end{equation}
Specifically, the distribution $p(k)$ applies to a given $z$ and $y$. The $p_i$ are ordered as in (4) so that a $k$ value that leads to the value $p_1(K)$ is a most likely estimate of $K$ from Eve's given $z$ and $y$. 

Note that it is this whole probability profile $p(k|y,z)$ that represents the general results of Eve's attack, which \textit{are not} simply an estimate of $K$, and as we will see, the results cannot be replaced by a single numerical criterion. In a classical situation, $p(k|y,z)$ is obtained from the ``channel" transition probability $p(y|x')$ and the side information $z$. In the quantum case, there would be infinitely many such transition probabilities, depending on what quantum probe and what quantum measurement Eve chooses to make. A security proof has to address all such possible $p(k)$ under a specific class of attacks or all possible attacks allowed by the law of physics, as in condition (i) of unconditional security.

It has been explicitly shown in Section III that an information-theoretic single-number security criterion merely puts a constraint on what possible $p(k)$ Eve may obtain, and it must satisfy the criterion constraint. For mutual information, the constraint states that $p(k|y,z)$ must not give a higher $I_E$ value that is ruled out by the security proof. Under the $I_E$ criterion, $p_1(K)$ given by Eqs. (9) and (10) shows that $I_E \to 0$ exponentially in $|K|$ does not imply that $K$ provides good security asymptotically. Here, we ignore the random variations in all parameters except $k$ by focusing on $p(k)$. Security will be weakened when these random variations are considered in Sections V and VIII.

Classically, it is already more convenient theoretically to measure the imperfection of $K$ by its statistical distance (variational distance [14], Kolmogorov distance) $\delta(K,U)$ from the uniform distribution $U$ than by $I_E$ as follows. For two probability distributions $P$ and $Q$ on the same sample space, the $\delta(P,Q)$ is defined to be

\begin{equation}
\delta(P,Q) \equiv \dfrac{1}{2} \displaystyle\sum\limits_{i} {|P_i-Q_i|}
\end{equation}
Thus,  $0 \leq \delta(P,Q) \leq 1$. From the well-known inequality in [14, Eq. (11.137)], it immediately follows that, for any subsequence or ``segment" $K^*$ of $K$ and denoting $\delta(K,U)$ by $\delta_E$ similar to $I_E$,

\begin{equation}
p_1(K^*) \leq 2^{-|K^*|} + \delta_E
\end{equation}
The result in [14, Eq. (11.137)] applies to any probability value $p(k^*)$ and hence to the maximum $p_1(k^*)$ in particular. Under the constraint $\delta_E \leq \epsilon$, it is easy to show by explicit construction [5] that the bound (13) can be achieved by many permissible $p(k)$. (We will often omit the unnecessary $d \leq \epsilon$ or $\delta_E \leq \epsilon$ in the following and simply set a $d$ or $\delta_E$ level.) In particular, for any $K^*$, one can achieve the bound (13) with equality. The case of the entire key $K^*=K$ for the total compromise probability

\begin{equation}
p_1(K) \leq 1/N + \delta_E
\end{equation}
is of special importance. These classical results directly apply to the quantum case in which a quantum trace distance $​d(K)$ ​is defined between Eve's probe and an ideal quantum state to the users. After Eve measures on her probe, a trace distance bound on Eve's attack simply translates to a bound on $\delta(K,U)$.​ ​Such a bound constrains the $p(k)$ that Eve could obtain from any probe and quantum measurement. Thus, in this paper, one can \textit{regard} $\delta_E(K)$ as $d(K)$ in the context of a quantum protocol.

Note that (14) shows the \textit{e​xponential ​p​roblem} in numerical security guarantee through $\delta_E$ similar to (10) above. An exponentially small $\delta_E=2^{-​​l}$ only gives security on $p_1(K)$ corresponding to an $l$-bit uniform key after dropping the small $1/N$ factor in (14). In particular, achieving (14) with equality already \textit{shows} that the failure probability interpretation of $\delta_E$ is logically incorrect because it does not include the $\nicefrac{1}{N}$ factor. 

\subsection{Failure Probability and Failure Probability Per Bit}

The original ​interpretation of the quantum trace distance $d(K)$​, which may be abbreviated as $d$, is based on a​ ``failure probability" interpretation ​on the classical $\delta_E$, which $​d$ would reduce to upon Eve's​ measurement ​​on her probe. A key $K$ is called ``$\epsilon$-secure" when $d \leq \epsilon$. ​It is stated in [30, p.414] that ``an $\epsilon$-secure key can be considered \textit{identical} to an \textit{ideal} (perfect) key- except with probability $\epsilon$" (emphasis in original statement). In addition, in [29, p.414], it is stated that ``the real and the ideal setting can be considered identical with probability at least $1-\epsilon$ ". Therefore, the ``failure" in the ``failure probability" refers to $K$ being not perfect, and a \textit{failure probability} $P_f \leq \epsilon$ guarantee means that it is rigorously proved that there is a maximum probability $\epsilon$ that $K$ fails to be perfect. This unambiguous and incorrect interpretation is repeated in many papers; see note [25] of ref [8] for a collection of cases. This is also explicitly asserted in the review [2] and in the most complete QKD security proof available to date [36]. 

This error has never been acknowledged, and the failure probability interpretation is widely perceived to be correct. The valid consequence, Eq. (14), of a $d$ guarantee is stated explicitly in [35]; however, the incorrect interpretation is maintained as a vague paraphrasing without noting the difference, and an indistinguishability argument is offered for such an interpretation. The security consequences of an incorrect interpretation will be presented in Sections V to VIII. Removing such a misinterpretation is important to obtaining true and proven security. In Section IV.A, we will analyze the errors committed in drawing the failure probability interpretation. In Section IV.B, we will do the same for the ``indistinguishability" argument, which is often taken to imply the same incorrect interpretation. In Sections IV-VIII, we will see in many places how the wrong interpretation misrepresents the security situation, attributing a security guarantee to $d$ that it does not provide.

The above failure probability claim was drawn​ from Prop. 2.1.1 in [29], which is the same as Lemma 1 in [30]. It is re-stated as Theorem A.6 in [35]. The claim states that, for two random variables $X$ and $Y$ in the same space with probability distributions $P_X$ and $P_Y$ that are marginals of a joint distribution $P_{XY}$, one may obtain

\begin{equation}
P(X=Y) = 1 - \delta(P_X,P_Y)
\end{equation}
Generally, for arbitrary $P_{XY}$, one obtains the following ``coupling inequality" [37, Sections I.2 and I.5],

\begin{equation}
P(X=Y) \leq 1 - \delta(P_X, P_Y)  
\end{equation}
Thus, (15) amounts to the assertion that the bound (16) can be achieved by some $P_{XY}$. Applying (15) with $X=K$ and $Y=U$, the probability $P(K=U)$ is taken to be a probability that $K$ is not $U$, and the failure probability interpretation of $\delta_E$ was thus drawn.

This is an example of interpreting mathematical symbols incorrectly when addressing real-world applications. There are several other such examples of incorrectly connecting mathematics and the real world in QKD security analysis, say, in connection with ``indistinguishability" and ``universal composition", as we will see later. The symbol $P(X=Y)$ merely abbreviates the probability of an event in which the outcome of $X$ equals the outcome of $Y$ from an applicable joint distribution $P_{XY}$, namely,

\begin{equation}
P(X=Y) = \displaystyle \sum\limits_{i} {P(X_i =Y_i)} 
\end{equation}
This does not say anything about the whole $X$ and $Y$ themselves, as the failure probability interpretation claims. More importantly, there is no joint distribution at play in this QKD situation other than the independent product distribution $P_X \cdot P_Y$, much less one that achieves the bound (16). The incorrect failure probability interpretation reads into (15) meaning which is not warranted. That it is wrong can be observed directly [5] from a $p(k)$ that achieves the bound (14), which has the additional factor $1/N$ exceeding what is given by the failure probability interpretation. When $\delta_E > 0$, the two distributions are necessarily different because $\delta(P_X,P_Y) = 0$ if and only if $P_X=P_Y$. In what sense then can $K=U$ hold with a probability when $\delta_E>0$?

Such a probabilistic interpretation for given $P_X$ and $P_Y$ may be represented mathematically by the existence of a distribution $P'$ such that, from the theorem of total probability,

\begin{equation}
P_X =  (1-\lambda)P_Y + \lambda P'
\end{equation}
for a probability $\lambda$, in this case, $\lambda=\delta(P_X,P_Y)$. Because $P'$ is a probability distribution, Eq. (18) is easily shown [38] to hold if and only if, for $X = K$ and $Y=U$,

\begin{equation}
(1 - \lambda)/N \leq p_i \leq \lambda + (1-\lambda)/N 
\end{equation}
For large $N$, up to which $i$ varies, Eq. (19) implies all $p_i$ take essentially the same value of approximately $\lambda$, and hence, $P$ must be nearly uniform. This condition (19) cannot be satisfied for $\lambda=\delta(K, U)$ [8]. For any $\lambda$, Eq. (19) implies a uniformity on $p_i(K)$ that does not follow from simply a guarantee on $\delta_E$. Specific counter examples can be easily constructed.

Thus, several errors are committed in the original derivation of the incorrect failure probability interpretation, any of which would invalidate the derivation. We omit a detailed discussion on the first two points, which are rather self evident.

\begin{enumerate}[label=(\roman*)]
 \item There is no reason to expect that maximizing $P_{XY}$ is in effect so that Eq. (15) holds despite (16).
 \item The mathematical representation of the failure probability interpretation of $\delta_E$ is not given via a joint distribution $P_{KU}$, which is irrelevant to such an interpretation.
 \item The correct representation of the failure probability interpretation is given by Eq. (18), which cannot hold for $\lambda=\delta_E$ and which is also not warranted for any $\lambda$ because of (19).
\end{enumerate}

Note that $d(K)$ from (14) gives the \textit{``total compromise probability"} $p_1(K)$ of the whole $K$ associated with $d(K)$, which is \textit{not} the probability $P_f$ that $K$ turns out to be non-uniform, apart from the $\nicefrac{1}{N}$ factor. This is because $P_f \leq \epsilon$ implies $p_1(K) \leq \epsilon$ but not the other way around, as we have shown.

A quantum trace distance measure $d/|K|$ obtained by dividing $d$ by $|K|$ and called the \textit{failure probability per bit} is introduced in [35], which clearly gives a substantially lower failure rate than does $d$ itself. It is a misleading terminology because it suggests that the bits in $K$ are statistically independent. With such an interpretation, the total compromise probability $p_1(K)$ becomes not (14) but the following $P_f$:
\begin{equation}
p_1(K) \sim\ d \text{ } \text{ } \text{ versus } \text{ } \text{ } P_f = (\dfrac{d}{|K|})^{|K|}
\end{equation} 

Two errors are committed in $P_f$ above obtained from a given $d(K)$. Instead of applying $d(K) \sim p_1(K)$ to $K$ as a whole, it is arbitrarily reduced by $\nicefrac{1}{|K|}$ to give a ``per bit" level $\nicefrac{d(K)}{|K|}$ \textit{and} is then applied to each bit of $K$ independently regardless of the length $|K|$.  As a result, the $p_1(K)$ level is greatly underestimated as the $P_f$ in (20). Generally, dividing a quantity such as $I_E$ and $\delta_E$ by the size $|K|$ does not produce a bit-independent quantity, as a matter of course.

This incorrect interpretation of ``failure probability per bit" is used in [35, p.14]:

\begin{enumerate}[label=(F)]
\item ``For example, if an implementation of a QKD protocol produces a key at a rate of 1 Mbit/s with a failure per bit of $10^{-24}$, then this protocol can be run for the age of the universe and still have an accumulated failure strictly less than 1."
\end{enumerate} The failure probability per bit here is $\nicefrac{d}{|K|}$ with the $P_f$ of (20). The numerical security situation of (F) is given in Section VIII.B. 

The failure probability per bit interpretation misses the crucial point that the significance of a given $d(k)$ level depends strongly on $|K|$. A level of $2^{-10}$ may be good for $|K| = 1$ but is poor for $|K| = 10^3$. On the other hand, the guaranteed level (13) gives the same bound $d(K)$ on the difference from a uniform distribution value $2^{-|K^*|}$ independently of the length $|K^*|$. Thus, a $d(K)$ value that appears to be small may actually be relatively large for a long $K$ or subsequence $K^*$. This confusion occurs in the same manner in the following distinguishability advantage interpretation of $d$.

\subsection{Distinguishably Advantage}

The indistinguishability argument was ​originally used in [28] and previously to argue that the trace distance $d$ in the quantum case or the statistical distance $\delta_E$ in the classical case is a good measure of how close $K$ is to an ideal situation for the users. It is precisely formulated as a distinguishability advantage statement for the binary decision problem of discriminating between the two distributions for $K$ and $U$. We will simply consider the $\delta_E$ case because the quantum detection problem reduces to a classical one once the (optimal) quantum measurement is fixed.

For the distinguishability advantage interpretation to serve as a functional security criterion, say, on KPA, one must first write down what the interpretation asserts quantitatively for a given problem. It appears, as this section will describe in detail, that the incorrect failure probability interpretation of Section IV.A is being asserted. Specifically, distinguishability is supposed to provide a justification of the interpretation. In addition to the counter examples of Sections IV.A and V.A to the failure probability interpretation of $d$, we show in this section how such a justification is conceptually invalid.

Consider the well-known classical binary decision problem of deciding between two hypotheses H\textsubscript{0} and H\textsubscript{1} from an observed random variable with conditional distribution $P_0$ and $P_1$ for the two hypotheses. The maximum probability of a correct decision $P_c$ is given by

\begin{equation}
P_c = \dfrac{1}{2} + \dfrac{1}{2}\delta(p_0P_0, p_1P_1)
\end{equation}
where $p_0$ and $p_1$ are the \textit{a priori} probabilities of H\textsubscript{0} and H\textsubscript{1}. In (21), the $\delta(.,.)$ is defined exactly as in (12), with $p_0P_0(i)$ and $p_1P_1(i)$ taking the place of $P_i$ and $Q_i$. When $p_0=p_1=1/2$, the second term on the right-hand side of (21) becomes $\dfrac{1}{2}\delta(P_0,P_1)$ in terms of an usual statistical distance. For this equal \textit{a priori} probability case, $\delta_E$ becomes the ``distinguishability advantage" of knowing $p(k)$ compared to the no observation case with the a posteriori probabilities of H\textsubscript{0} and H\textsubscript{1} equal to the \textit{a priori} probability $1/2$. Thus, it is thought that if $\delta_E$ is small, $K$ is hardly ``distinguishable" from $U$.

It is easily observed from (21) that $P_c$ is biased toward hypothesis H\textsubscript{0} when $p_0>p_1$ and similarly for H\textsubscript{1}. When $p_0$ goes to 1, $P_c$ for H\textsubscript{0} also goes to 1, as it should intuitively.  It is not known how (21) is related to the distinguishability advantage $P_c-p_0$ when $p_0>1/2$. Thus, as it can already been observed from simply the problem formulation, the criterion $\delta_E$ can only admit a distinguishability advantage interpretation in cryptography, if at all, for $p(\text{ideal})=p(\text{real})=1/2$, with $p(\text{ideal})$ being the \textit{a priori} probability $p_0$ of the hypothesis H\textsubscript{0} that the situation is perfect for the users and $p(\text{real})$ for the actual situation where Eve has made an observation described by her $p(k)$. It is our contention that this requirement of $p(\text{ideal})=1/2$ is not realistically meaningful, and furthermore, \textit{no} quantitative security conclusion on Eve's success probabilities can be drawn from such interpretation; in particular, the bound (13) or (14) must be derived mathematically from the mathematical expression of $\delta_E$ with \textit{no} extraneous interpretation. The following already presents that the conclusion of the real situation being ``indistinguishable" from an ideal one or having ``distinguishability advantage $\delta_E$" cannot be validly drawn. The conditional probability whereby the situation is ideal for the $p_0=1/2$ case from binary discrimination is given by\begin{equation}
\begin{split}
P(\text{ideal}|\text{H}_0 = \text{ideal}) = 1/2 + \delta_E/2, \\
P(\text{ideal}|\text{H}_1=\text{real}) = 1/2 - \delta_E/2 
\end{split}
\end{equation}
Why would the ideal situation have such a high probability close to $1/2$ for any $\delta_E \ll 1$? This is because the \textit{a priori} probability $p(\text{ideal})$ is taken to be 1/2 to begin with even though it should be 0. 

Indeed, if it makes sense to assign an \textit{a priori} probability to the real and ideal situations in a binary discrimination problem, the \textit{a priori} probability of the real situation should be $1$, and the ideal situation should be $0$. This is also the conclusion drawn from (18)-(19) above. However, what if one simply ponders the decision problem with $p_0 = p_1 = \nicefrac{1}{2}$? Then, the conclusion of such a problem cannot be applied to any real-world problem. This is because such a discrimination problem has no empirical meaning because we all know we are in the real situation where Eve's presence is assured. If this may not be the case, the problem should be formulated as one with all possible unknown probabilities of Eve's absence or ``false alarms" [39] and not one with a fixed \textit{a priori} probability.

Furthermore, Eve never cares to make such a discrimination; her objective is to learn about $K$. This is another case of reading into mathematics an unwarranted assertion about the real world. We will elaborate this point further in the remainder of this section because there is a similar use of $\delta_E$ in conventional cryptography that we cannot go into in this paper and that lends unwarranted security significance to ``indistinguishability". (In particular, correlations between the future bits are not accounted for in the single-bit ``distinguisher" prediction, similar to what we indicated above at the end of Subsection IV.A.)

The problem of a ``metaphysical" distinguishability interpretation can be observed from the fact that there are \textit{many} hypothetical situations, say, one with any  $\delta_E$ level, in addition to the $\delta_E=0$ ideal case. Should we conduct multiple-hypothesis decision making? Why not a binary one with one situation less secure than the real one? Why would such a decision allow one to conclude all the features of the decided upon hypothesis, which are simply given by fiat?

One major problem of using such an distinguishability advantage argument is that it \textit{becomes} in one's mind an \textit{indistinguishability} statement when the distinguishability advantage $\delta_E$ is small. Indeed, the real situation and the ideal situation are then taken to be distinguishable and hence different only with probability $\delta_E$. Thus, $\delta_E$ becomes the failure probability interpretation of the previous section! Such an explicit interpretation of quantitative indistinguishability as failure probability is common; see for example [40, p.3]. Moreover, it appears to be used by many as a valid derivation of the failure probability interpretation of $\delta_E$ despite the errors in the original derivation discussed in Section IV.A and the abundance of counter-examples to such a claim, the reasons for its invalidity notwithstanding. The following may help further clarify what went wrong.

There is a common-sense meaning of two items being ``indistinguishable," with the Leibniz metaphysical principle concerning the ``identity of indiscernibles" implicitly used. That such an indistinguishability conclusion cannot be drawn from a binary decision problem can be directly observed from the common problem of radar detection as to whether there is an incoming flying object. In a militaristic situation, the object of concern could be an enemy airplane, say, with or without a warhead. The yes-no target detection problem of whether an enemy airplane is present \textit{cannot} alone determine whether a warhead is on board. In the absence of further information, one cannot infer that the airplane has a warhead because that is hypothesis H\textsubscript{1} in the binary decision problem formulation, which one simply applies \textit{by hand}. There are many possibilities on the details of the incoming target; it is not valid to pick one and exclude others and then use binary discrimination to affirm the picked possibility. Similarly, the occurrence of an ideal situation is an unwarranted conclusion that one cannot make use of in other problems. One has to \textit{mathematically derive} a result for a problem from the given mathematical statement $\delta_E \leq \epsilon$. 

Indistinguishability arguments in terms of $\delta_E$ are supposed to be ``universally composable" in that they justify the use of $K$ in any application to which it would be applied [28], [29]. We would later run into such issues in connection with known-plaintext attacks and error correction. Here, we may simply emphasize that there is no such automatic universal composability from $\delta_E$ or $d$, with or without a  distinguishability advantage. Even in the case whereby it is composable, the proof from $d$ may be far from trivial, as we will also observe. The main point is that an intuitive interpretation may only serve as a guide to the general situation. Valid logical and mathematical deduction from premise to conclusion is required to establish proof. This is especially clear when a quantitative level is desired.

Summarizing, the ``distinguishability advantage" justification of operational guarantees from $\delta_E$ or $d$ is incorrect in several ways:

\begin{enumerate}[label=(\roman*)]
 \item The indistinguishable probability is for a specific binary decision problem, which does not imply that the two situations are indistinguishable in other physical senses. 
 \item In the real world, the \textit{a priori} probability for the ideal situation cannot be $1/2$; instead, it should be $0$.
 \item The ideal situation cannot be inferred to be the real situation from the binary decision because it includes other features not included in the binary hypothesis testing formulation.
\end{enumerate}

The key question to ask concerning the ``distinguishability advantage" argument is what is the quantitative security assertion? It seems that the answer so far is the failure probability interpretation, to which we have given various counter-examples in Section IV.A and will give another in Section V. 

\section{Some Correct Guarantees And Open Problems With The Criterion $d$}

\subsection{Guarantee On Known-Plaintext Attack}

The bound (13) provides a security guarantee on the security of $K$ and its subsequences $K^*$ when Eve attacks $K$ during its generation process before it is used. As mentioned before, when $K$ is used for privacy, it is more important to protect against known-plaintext attacks to maintain the secrecy of $K_2$ when $K_1$, namely, the remainder of $K$, is known to Eve. How does such security follow from the incorrect failure probability interpretation of $d$?

It seems that this issue is addressed explicitly only in [35, Section 5.1], with the conclusion that the security is the same as that originally obtained from $d$. Indeed, the failure probability interpretation alone without the indistinguishability and other considerations in [35] would appear to give such a conclusion already. Thus, regardless of the $K_1$ known to Eve, $K_2$ remains uniform to her except for a probability $d$. Note that the distinguishability interpretation gives exactly the same quantitative security conclusion as the failure probability interpretation, as noted in Section IV.B.

Such a conclusion is​ ​\textit{incorrect}, as the following counter-example shows. Let $k^0$ be a specific sequence of $k$ with probability $p(k^0) = 2^{-m}$. We denote the first $m$ bits of $k^0$ by $k_1^0$ and the remaining $n-m$ bits by $k_2^0$. Let all the other sequences $k$ with the first $m$ bits $k_1^0$ have $p(k) = 0$. The other sequences $k$ with the first $m$ bits different from $k_1^0$ are assigned a probability $2^{-n}$ as in $U$. Then, $\delta_E = 2^{-m}$. When the known $k_1$ in a KPA turns out to be $k_1^0$, Eve knows that the remainder $k_2$ is $k_2^0$ with certainty, \textit{not} with probability $d$.

The underlying reason ​that the failure probability cannot be used to obtain correct results in KPA ​is that there is no way to account for \textit{conditioning} ​with just such an interpretation​, and ``universal composition" is a vague argument and not universal. Its validity needs to be established for each composition situation. In KPA, there is apparently no composition, and thus, the original $d$ result is inferred in [35] as noted above, which is numerically very incorrect.

Apparently, KPA security on $p_1(K_2^*|K_1)$ of (6) can be obtained directly from (13) by writing its left-hand side as an average, with $K_2'$ being the remainder of $K_2$ apart from $K_2^*$:
\begin{equation}
\sum\limits_{k_1} p_1(K_2^*|k_1)p(k_1) \leq 2^{-|K_2^*|} + \delta_E
\end{equation}    
for
\begin{equation}
p_1(K_2^*|k_1) = \sum\limits_{k_2'} p_1(K_2^*|k_1,k_2')p(k_2'|k_1)
\end{equation}
It is observed from (23) that the probability guarantee for KPA now is an average over $K_1$. This is fully in accord with the above counter-example. When we remove the $K_1$ average to obtain an individual guarantee, we need to apply a Markov inequality for $K_1$. The drawing of a specific $d$ from possible PACs and a specific $y$ from an observation $Y$ also requires a Markov inequality. This will be discussed in Section V.C.​ In Section V.D, we will make clear that all these classical results are what quantum results reduce to. ​

\subsection{Bit Error Rate (BER) Guarantee​}

The following security question arises: how many bits will Eve correctly obtain even though her estimate of $K$ or $K_2^*$ is incorrect as a sequence under a $d$ guarantee through (13) or (23)? For example, Eve guessing a 4-bit $K$ to be $0010$ when it is actually $0011$ makes a sequence estimate error but only a one-bit error out of four, not the uniformly random result of two errors. However, an error rate leak in security that differs from the uniformly random level of $\nicefrac{1}{2}$ is equivalent to a non-uniform \textit{a priori} distribution $p_0(K)$, which is known to Eve. For instance, Eve knowing that six out of ten bits of a $K^*$ are correct but not knowing which are the correct bits is equivalent to having an \textit{a priori} $p_0(K^*)$  with a biased probability $0.6$ on each single bit in the case that the bits are independent. Hence, the issue must be addressed in assessing ultimate security. 

In ordinary communications, this is called the \textit{bit error rate} (BER) issue in coded systems, in contrast to the sequence error rate addressed in most performance analyses. This does not represent a serious issue there because, typically, a sequence error rate itself can already be driven to zero and because BER is in any case an improvement over the sequence error rate. The main information-theoretic problem to cryptography users concerns Eve's performance from the users' viewpoint, which is \textit{opposite} to the performance concern of the users themselves. It turns out that the relative importance of the different issues may be different in cryptography in addition to the fact that the required performance analysis is often more difficult, say, in lower bounding instead of upper bounding Eve's error probability. 

The BER is $1/2$ when $K=U$ under any attack.​ From the failure probability interpretation of $d$, one would obtain, for any subset $K^*$ of $K$ in the absence of known-plaintext attacks, the following bound on such a bit error rate.

\begin{equation}
\text{BER} \geq (1-d)/2
\end{equation}               
Counter-examples to (25) can be ​read​​ily constructed​​ for small $n$​.​ The actual BER needs to be validly bounded from a given $\delta_E$. Eve's BER, which is less than $\nicefrac{1}{2}$, gives her information that is not available for a perfect $K=U$, and its quantitative security consequence needs to be obtained. 

It is important to observe that the BER is a very important security criterion in addition to those of (6); however, it alone is not sufficient as a security guarantee. All these different probability criteria arise naturally for different security concerns for a given $p(k)$, and all have direct operational meaning. They are perfectly protected against Eve when $K = U$.

An approximate ​bound for t​he whole $K$ can be deriv​ed [41] from standard information theory results through the entropy $H(K)$ of $K$. Let the bit error probability be
\begin{equation}
p_b \equiv P_b(k) = \dfrac{1}{n} \sum\limits_{i}P_e(i)
\end{equation}   
where $P_e(i)$ is the probability that the $i$th bit in $K$ is incorrectly obtained from her regardless of the estimate of $K$. With $H_2(\cdot)$ being the binary entropy function, we have​ from the Fano inequality [8]​ that
\begin{equation}
nH_2(p_b) \geq H(K) - I_E
\end{equation}                   
The right-hand side of (27) can be bounded via $\delta_E$ for $\delta_E \leq \epsilon$ ​by neglecting $I_E$ compared to $H(K)$ and using the theorem in [14, p.664],​
\begin{equation}
H(K)​ \text{ } \utilde{>} \text{ }​ n - 2\epsilon(n + \log{\dfrac{1}{2\epsilon}})
\end{equation}                    
A bound on $p_b$ follows from combining Eqs. (27)-(28); see [41] for a discussion on the relatively weak bound on such $p_b$ in comparison to sequence errors. Note that the bound (13) for a single-bit $K^*$ does not concern the BER, which is obtained from a sequence estimate of a long $K^*$ with some bits being correct even though the sequence is wrong.

For a general subsequence $K^*$, there is no known result on the BER guarantee from $\delta_E$ or $I_E$ and none for $K_2^*$ under KPA with known $k_1$.​ The bound (25) from the failure probability interpretation ​on the BER for KPA is contradicted by the same counter-example in Section V.A​.​ Summarizing, it is uncertain as to what BER guarantee for Eve one can have under $d \leq \epsilon$.​ Useful bounds on Eve's BER for these cases are \textit{open} problems with basic security significance, as we will see in connection with error correction in Section VI.

\subsection{Necessity Of Individual Guarantee And Consequences}

In the industrial control of product manufacturing, the criterion employed is usually the probability an item fails to meet a pre-set standard, not the average number of failures. The former is a more stringent and operationally meaningful criterion, as we will see. If we have a zero-one random variable, then the probability of one variable is the same as the average. Otherwise, from the average $E[Z]$ of a positive-valued random variable $Z$, we can bound the probability that $Z$ exceeds a given level $\gamma > 0$ by the Markov inequality [14]

\begin{equation}
P_r[Z \geq \gamma] \leq E[Z]/\gamma 
\end{equation}
The Markov inequality is needed because, often, only $E[Z]$ may be evaluated or bounded and because $P[Z \geq \gamma]$ cannot be obtained via another route. This is the situation in the QKD security proofs.   

There are at least three reasons why probability guarantees should be used instead of average guarantees. First of all, consider the (artificial) example of a cryptosystem that has a 50/50 chance of being secure for 100 years or totally insecure. Its average security duration is 50 years; however, no one would find such security meaningful. Second, the average has no direct operational meaning when the trial sample size is small relative to the size of the probability (sample) space for a single trial. This is evident in the case of one sample trial, the meaning of probability for which has been extensively discussed; see, for example, [42]. There are insufficient samples for the average effect to kick in with a small variance in the sample mean (average), and we should use probability to assess​ the ​trial sample property. This is found to always be the case in QKD protocols. A third reason is that an empirical average (sample mean) is not a deterministic quantity. An empirical average lacks operational significance without some guarantee that the spread around the mean is sufficiently small.

In QKD protocols, there are many random parameters with a probability distribution. The final trace distance $d$ obtained is itself an average over the possible PAC codes, only one of which is used in a single round (see Section VI.A). ​Such an average result is common in information theory work with the so-called ``random coding" argument. ​The possible number of ​PAC ​codes is substantially larger than $2^{|K|}$ even if a Toeplitz matrix is used as the PAC, as is commonly the case. Therefore, the averaged $d$ has to be first converted into an individual $d$ from (29). Then, a measurement of some $Y$ is involved in Eve's attack, which is the suppressed $y$ dependence of Eve's $p(k)$ that we focused on previously. Either classically or quantum mechanically, the guarantee is on the average over $Y$, and it also needs to be converted into a probability​ for a specific $y$​. The number of possible $y$s again far exceeds $2^{|K|}$. These two averages are both on $d$ and can be combined. There is another​ average over the known $K_1$ in KPA that we need to address in the case of protection against known-plaintext attacks discussed in Section V.A. Therefore, in total, we would need to apply (29) one or two times.

The inequality (29) allows one to exchange the level of the failure threshold $\gamma$ with another ``failure probability" level $P_r[Z \geq \gamma]$ that exceeds the desired threshold level. In QKD security proofs, the $Z$ itself is a probability, as observed in Sections IV and V, and is a random quantity depending on the values of several other random system parameters. If we did not use these two probabilities, we would not know quantitatively what individual probability guarantee we may obtain. We could address such uses of (29) in the final security guarantee by adjusting the exchange to minimize any ``failure probability" of the protocol as follows. 

Consider the $p_1(K^*)$ of (13) when $2^{-|K^*|}$ is negligible compared to $\delta_E$. Then, with $p_1(d\geq\delta)$ denoting the conditional probability of $p_1(K^*)$  given $d\geq\delta$, etc., we have, under $E[d] = \epsilon$ and $\delta = \epsilon^{\nicefrac{1}{2}}$,

\begin{equation}
\begin{split}
p_1(K^*) &= p_1(d\geq\delta) P(d\geq\delta)+p_1(d<\delta)P(d<\delta)\\ &< 1 \cdot \dfrac{\epsilon}{\delta} + \delta\cdot 1 = 2\epsilon^{1/2}
\end{split}
\end{equation}
We have used (29) and minimization over $\delta$ to arrive at the guarantee (30). Similarly, abbreviating $p_1(K_2^*|k_1)$ by $p_1$ with $2^{-|K_2^*|}\ll f$ for $E_{K_1}[p_1] = f$, we have
\begin{equation}
p_1 = p_1(d\geq f)P(d\geq f)+p_1(d<f)P(d<f)
\end{equation}
\begin{equation}
\begin{split}
p_1(d<f) &= p_1(d<f,p_1\geq g)P(p_1\geq g)\\
&+ p_1(d<f,p_1<g)P(p_1<g)
\end{split}
\end{equation}
Thus, from (29) and minimization over (f,g),

\begin{equation}
\begin{split}
p_1<1\cdot \dfrac{\epsilon}{f}+1\cdot \dfrac{f}{g}+1\cdot g = &\dfrac{\epsilon}{f}+\dfrac{f}{g}+g \\< 3\epsilon^{1/3} \text{ for } f = \epsilon^{2/3}=g&^2
\end{split}
\end{equation}
The bounds (30) and (33) are loose; however, it appears that there is no way to tighten them without further knowledge of the random system parameters.

Because the BER is a very nonlinear function of the security criterion $\delta_E$, it is not known how the average $\delta_E$ can be converted into an individual guarantee, in contrast to the $p_1(K^*)$ or $p_1(K_2^*|K_1=k)$ case above. Of course, we do not even have a BER bound without such an average, except for the whole $K$ from (27)-(28).

The numerical security guarantee from (30) and (33) is devastatingly worse than the original $\epsilon$-level guarantee. Indeed, even with the incorrect failure probability interpretation of $d$ discussed in Section IV.A, one application of (29) is required to obtain an individual guarantee on Eve's probability of successfully estimating the whole $K$​ even without any side information on $K_1$ during its use, as discussed above; see Section VIII for numerical examples.

\subsection{Validity of Classical Information-Theoretic Results in QKD}

At this point, it is appropriate to emphasize that the classical analysis of $\delta_E$ and $I_E$ that we presented in this paper applies directly to the quantum case. This is because, regardless of the utilized quantum criterion, $d$ or otherwise, the criterion would reduce to a classical quantity once Eve makes her quantum measurement on her quantum probe. The trace distance $d$ would reduce to a classical statistical distance $\delta_E$, and the accessible information would reduce to classical mutual information. However, different quantum quantities may lead to the same classical quantity but are essentially different quantum mechanically. This turns out to be the case for quantum accessible information and the Holevo quantity; a guarantee from the former allows quantum information locking leaks, which is not the case for the latter [43]. The  Holevo quantity guarantee is essentially equivalent to the trace distance $d$ guarantee. From this quantum equivalence [28], [43], one  immediately has the following bounds, which establish the essential equivalence of $\delta_E$ and $I_E$ in a classical protocol and which provide a general security guarantee to classical protocol security proofs via $I_E$ similar to that provided by $\delta_E$ given in this paper:
\begin{equation}
2{\delta_E}^2 \leq I_E \leq 8n\delta_E + 2H_2(2\delta_E)
\end{equation}
​The $\delta_E$ in (31) is an average over the observation $Y$ in classical protocols, which is automatically included in the quantum trace distance. This is exactly as in the $I(K;Y)$ case.​

\section{Information Leak From Error Correction and Privacy Amplification}

In this section, we consider the problem of quantifying the security of the ECC output $K'$ and the PAC output $K$, the generated key in Fig. 2, as well as how ECC and PAC affect the final key generation rate. The information leaks from error correction and privacy amplification were not considered in the earlier security proofs [7], [26], [27]. This is sometimes justified by the invalid reason that the open exchange in these two steps is performed after Eve sets her quantum probe. However, Eve may make her quantum measurement and key estimate after the open exchange. Apparently, the PAC step can be rigorously quantified if the ECC step has also been quantified; however, the ECC step cannot be quantified, and there is no hint as to how a rigorous quantification of error correction may be performed in a QKD protocol. Before discussing the error correction problem, we first discuss privacy amplification and its effect on the key rate.

\subsection{Privacy Amplification}

The basic idea of privacy amplification is to increase the security level by compressing the input bit sequence into a shorter output bit sequence. Intuitively, this is well known to be possible when the input bits are statistically independent to Eve. For example, given two bits $x_1$ and $x_2$, each known to Eve with error probability $p < \nicefrac{1}{2}$, the bit $x_1 \oplus x_2$ is known to her with error probability $1-2p+2p^2$, which is larger than $p$. When the input bits are correlated, if simply from Eve's possible attack, the situation is far less simple. Useful results can be obtained using linear hashing compression represented by a $|K'|\times|K|$ matrix via the so-called Leftover Hash Lemma [44], which has a direct quantum generalization [45]. The Leftover Hash Lemma for ``universal hashing," which covers all PAC in use, provides the tradeoff between the $d$-level $d(K)$ of $K$ and its length $|K|$ by the following formula, with  $p_1(K') = 2^{-l(K')}$,
\begin{equation}
|K| \geq l(K')-2\log{\dfrac{1}{d(K)}} \equiv g(K,K')
\end{equation}
Because it is not known whether $|K|$ greater than the minimum $g(K,K')$ on the right-hand side of (2) could be obtained, the guaranteed key rate is given by the quantity $g(K,K')$. Note that $0 \leq d \leq 1$ and $|K| \leq l(K')$. Furthermore, the minimum $d(K)$ one can obtain is, from $0<g(K,K')$,
\begin{equation}
d(K) \geq p_1(K')^{1/2}
\end{equation}
We can simply take the quantum $d(K)$ in this paper to be the largest statistical distance $\delta(K,U)$ that Eve may obtain. This Leftover Hash Lemma guarantee is an average over the family of possible hash functions from which the PAC is drawn. The specific PAC used is openly announced, and the performance is an average over possible codes, which is common in ``random coding"-type arguments.

A specific $\delta_E(K)$ or $d(K)$ level has to be first guaranteed in the security analysis to remove the PAC averaging. There are evidently some PACs with poor security, say, whenever the PAC matrix is degenerate (rank less than $|K|$), for which a degeneracy of $m \leq |K|$ would leak $m$ Shannon bits with certainty. If such degeneracy is first tested, a daunting practical task given that $|K|$ is tens of thousands  and given that $|K'|$ is a multiple of $|K|$, the resulting family is not known to obey  the ``universal family" condition required for the proof of the Leftover Hash Lemma. A high-probability guarantee on an adequate specific $d$-level is therefore essential.

The following inequality evidently holds for the $K$'s in Fig. 2:
\begin{equation}
p_1(K'') \leq p_1(K') \leq p_1(K)
\end{equation}
The first inequality in (37) follows from Eve possibly possessing more knowledge from the error correction in estimating $K'$. The second inequality follows from privacy amplification being an open many-to-one transformation. As will be discussed in Section VI.B, the users could and indeed may have to cover the chosen ECC via shared secret bits; therefore, one would obtain $K'' = K'$ assuming that \textit{correctness} (Alice and Bob agree on the same $K$) is obtained with a sufficiently high probability. In such a case,
\begin{equation}
p_1(K'') = p_1(K')
\end{equation}
When the ECC is covered by an imperfect key, there is no known bound on $p_1(K')$, as will be observed in Section VI.B. Hence, there is also \textit{no} guarantee on the $d$-level of the final $K$ from (35), and the PAC step justification from (35) and (38) is lost because (38) is no longer valid. The rigorous validity of the final $d(K)$ level is correspondingly \textit{lost} from simply this problem.

Note that it is not possible to cover a PAC using shared secret bits because this would require a bit cost substantially greater than the number of key bits $|K|$ generated because $|K'|$ is typically many times $|K|$. The open announcement of PAC is fully considered in the Leftover Hash Lemma. There is no similar result that would yield the PAC information leak automatically from another known approach [46].

Privacy amplification exemplifies the exchange of key rate and privacy level inherent in QKD protocols. For PACs to which the Leftover Hash Lemma is applicable, there is the limit (35) on how small $d(K)$ can be made whereby $|K|$ remains positive. In general, from (35), $p_1(K'')$ sets a limit, via $p_1(U)=2^{-n}$, on the number of uniform key bits that can be generated, and $p_1(K'')$ is constrained by (3) in a $I_E$ guarantee and by (14) in a $\delta_E$ or $d$ guarantee on $K''$. Such an exchange is fundamental. It has not been shown how, and it appears impossible, one obtains a key at a given rate of $|K|$ per round with $K$ made arbitrarily close to perfect by increasing a security parameter in either a finite or an asymptotic protocol. In particular, it is not possible to obtain $p_1(K)$ arbitrarily close to $2^{-|K|}$ from repeated use of linear PAC, which is a direct consequence of (35)-(36). It is not known whether a PAC may exist that leads to a better exchange than (35). On the other hand, substantially more secure keys than those reported in the literature can be obtained from (35) at the expense of a decreased key rate [9]. In particular, a ``near-perfect" key $K$ with $d(K) = 2^{-|K|}$ may be obtained, although that alone does not address the unsolved security issues concerning the BER and ECC.

\subsection{Error Correction and a Main Unsolved Problem}

The error correction step is called ``reconciliation" in the early QKD literature and is to be achieved by an open exchange Cascade protocol [47]. There is no valid quantitative result on Cascade [48] because complicated nonlinearly random problems are involved. Furthermore, the difficulty of bounding the resulting $p_1(K')$ means that the subsequent PAC step cannot be quantified if one uses Cascade. The same situation is obtained when the error correction step is performed openly, as further discussed later in this subsection. 

Currently, ECC is universally employed for error correction in QKD protocols. In particular, large LDPC codes are used, the performance of which is difficult to analyze [44]. The problem of ECC information leaking to Eve was not mentioned in earlier security proofs [13], [26], [27], in [35] or in the recent review [2]; however, the added side information of an ECC on $K''$ would help Eve in her estimate $\hat{K}'$ of $K'$ if the ECC is openly known. In particular, if the ECC is too powerful, it may even correct all of Eve's errors in $K''$. As discussed in Section IV.A, for security quantification, one would need to bound $p_1(K')$, which is an impossible task even classically for any given long ECC. There is a further quantum issue [9] similar to quantum information locking concerning the accessible information criterion. Thus, the \textit{only} viable security approach is to cover the ECC using shared secret bits between uses and subtract its cost from $|K|$ to obtain the final generated key rate $|K_g|$. Indeed, the following formula is currently used:

\begin{equation}
{\rm leak}_{EC} = f \cdot |K''|\cdot H_2 {\rm (QBER)}
\end{equation}
with
\begin{equation}
|K_g| = |K| - {\rm leak}_{EC}
\end{equation}
The factor $f$ is arbitrarily taken to be $1 \leq f \leq 2$, the case $f = 1$ being the asymptotic $|K''| \to \infty$ limit. 

The justification of (39) is given in [36] by citing the whole book [14], which does not address such reconciliation issues or even ECCs. We give the following argument for the case $f=1$, which appears to be what is intended in the earlier paper [49].

Consider a linear $(m,k)$ ECC with $k$ information digits and $m$ code digits [50] such that the number of parity check digits is $m-k$. If one assumes that the $K''$ from the $X'$ transformation in Fig. 2 can be represented by a binary symmetric channel [8] with crossover probability given by the QBER, then for $k$ given by the channel capacity $1-H_2$(QBER), there exists a linear code that would correct the errors from the $|K''|$ received bits for large $|K''|$ by Shannon's Channel Coding Theorem, which is applicable to random coding over linear codes only. Hence, the number of parity-check bits that are to be covered by a one-time pad, with $m$ of the $(m,k)$ code being $|K''|$, is
\begin{equation}
|K''| - |K''|\cdot[1-H_2{\rm (QBER)}]=|K''|\cdot H_2{\rm (QBER)}
\end{equation}
Thus, (36) for $f=1$ is obtained.

We would first remark that the accounting in (39) regards the $K''$ sequence as a codeword of an ECC, which is sometimes explicitly stated in QKD security analysis. In such a situation, covering the parity check bits is not sufficient to uphold (38) needed for the PAC step. This is because the structural information on the specific ECC used, which is open because it would take an excessive number of shared secret bits to cover it, would induce correlations among the bits in $K'$ such that it becomes impossible to estimate the increase in $p_1(K'')$ to $p_1(K')$. Even the effective $K''$ itself has been changed when it is taken as a code word. On the other hand, by regarding $K''$ as the information digits of a linear ECC in a systematic form, Alice may simply add further parity check digits and cover them by a one-time pad, hence preserving (38). If the covered parity-check digits are assumed to be error free, then (39) continues to hold. In reality, the digits have to be error protected for the classical channel used for their transmission. If that channel is taken to have the same error rate give by QBER, a different ${\rm leak}_{EC}'$ is obtained because the $k$ of the $(m,k)$ code is now $K''$:
\begin{equation}
{\rm leak}_{EC}' = |K''| \cdot H_2{\rm (QBER)}/[1-H_2{\rm (QBER)}]
\end{equation}
which is larger than ${\rm leak}_{EC}$. The resulting $|K_g| = |K| - {\rm leak}_{EC}'$ will be correspondingly smaller. 

The combined key rate reduction effect of the PAC and only (39) is quite pronounced. In addition to the intrinsic physical inefficiency of QKD, they further severely limit the obtainable key rate in a full protocol.

There are several basic problems with such an approach to quantifying ECC security [9], [51]. The assumption of a binary symmetric channel is not valid under general attack by Eve; otherwise, there would have been no problem in quantifying the $K''$ security since QKD day one. The pulling back of the asymptotic $|K''| \to \infty$ limit to a finite $|K''|$ with an ad hoc factor $1 \leq f \leq 2$ is completely unjustified. Although the parity-check covering bit cost  in a concrete protocol  may be smaller than (40) when the protocol continues to function correctly (from other issues, such as correctness, that we do not discuss in this paper; however, see Section VI.C), the hand waiving assignment of $f = 1.1$ or $1.2$ in the literature shows that QKD security has \textit{not} been rigorously quantified in principle. This is because no correctness is guaranteed if an empirically measured quantity is used for the bit cost in lieu of (39). Some formal results on information leakage in open ECCs are presented in [52]; however, (39) is employed in actual evaluations [53]. 

Substantially more serious is the following basic security issue. The importance of QKD derives from the fact that key bits can be continuously generated between two users; in particular, such bits can be used to execute a future QKD protocol. Otherwise, one does not obtain effective key generation. We can perform the analysis above for ECC security only by assuming that the shared key bit used to cover the parity-check digits are the perfect one-time pad bits. When $K$ is not perfect, what would the information leak be? This problem is \textit{never} explicitly addressed in the literature. In the following, we will ascertain whether ``universal composition" may be of assistance.

Universal composition has been based on two different arguments. The standard one [28], [29] is the metric property of $\delta(P,Q)$ or of the quantum trace distance. For application to the present ECC problem, we have
\begin{equation}
\delta(P_\text{ideal},P_\text{ecc}) \leq \delta(P_\text{ideal},P_\text{no ecc}) + \delta(P_\text{no ecc},P_\text{ecc})
\end{equation}
where $P_\text{ideal}$, $P_\text{ecc}$, and $P_\text{no ecc}$ refer to Eve's distribution on $K$ for the ideal case $K=U$, the case when a specific ECC is used and the case when no ECC is used. In the quantum situation, the classical $\delta(P,Q)$ would be replaced by the trace distance between corresponding density operators, i.e., the quantum counterpart of classical distributions. From (43), we need to bound $\delta(P_\text{no ecc},P_\text{ecc})$ to obtain a $\delta_E$ level with ECCs, which appears to be an impossible task, and no result has ever been reported for carrying through this universal composition argument. There is no valid proof if $\delta(P_\text{no ecc},P_\text{ecc})$ is taken to be the $\delta_E$ level of the key $K_\text{ecc}$ used to cover the ECC. In particular, the BER leaks of $K_\text{ecc}$ discussed in Section V.B would alone give Eve significant side information to improve her estimate of $K_\text{ecc}$ and hence of $K''$. There is a complicated nonlinearity involved in these $\delta$ levels. 

The other argument [54] uses an incorrect and thus invalid failure probability interpretation of $\delta_E$ or $d$. If the argument were to be valid, then one would add the $d$-level of the ECC covering key to the overall $d$-level. As we have observed in Sections IV and V,  some details concerning $K$ are not protected by $d(K)$; however, they are protected under the incorrect failure probability interpretation. In particular, the interactions of the different parts of a protocol indicates that one may not need an entire portion to be correctly estimated to improve an overall estimate on another portion. How a BER leak of $K_{ecc}$ would affect Eve's success probabilities through the ECC appears to be a complicated function of the given ECC. There is no reason why $\delta(P_\text{no ecc}, P_\text{ecc})$ would equal $\delta(K_\text{ecc},U)$ in the absence of an explicit proof. Moreover, such a level cannot be generally correct because when the ECC is sufficiently powerful to correct all errors, the $m-k$ parity check digits would reveal all  $k$ bits of $K'$. However, the failure probability derivation of universal composition [54] needs such further proof. Using extraneous interpretation is not an alternative to a valid mathematical deduction.

The severity of the $d$-level limit on a QKD key in applications will be described in the following sections on message authentication. In the present ECC case, it appears extremely difficult to derive reliable estimates. One may thereby conclude that the security of the ECC step in a QKD protocol has not been, and appears that it cannot be, analyzed quantitatively in a valid manner. As a consequence, the PAC step is not justified due to the lack of a rigorous bound on $p_1(K')$, as discussed in Section VII.A. Hence, the security of the entire QKD protocol has not been reliably, and certainly not rigorously, quantified. This defect is not one of complexity in numerical evaluation but one of fundamental validity of reasoning.

\subsection{Fundamental  and  Practical  Limits  on  Key  Rate  and  Security  Level  Exchange}

In this section, we will summarize and explain the important basic and  practical limits on the exchange between $|K|$ and $d(K)$, including the possible adjustments of what $|K|$ is in a QKD round for such an exchange. Indeed, what constitutes a QKD round?

Let us first ignore the practical limits on processing long ECCs and PACs and simply attempt to determine what is a good choice of $|K''|$. Because $K''$ has to be error corrected, we need to introduce a measure of correctness, namely, the probability that the users agree on the same $K'$. In a realistic protocol, there are various system imperfections that would compromise correctness; however, the necessity of error correction alone implies that long ECCs need to be used. This arises from the fact that $K''$ being broken into small pieces for error correction is equivalent to using a shorter ECC in sequence as a longer ECC, which has never been found to be an efficient method of correcting errors. We simply have to use a sufficiently long ECC, or equivalently a sufficiently long $K''$, to achieve an adequate level of correctness, i.e., of correcting all  the errors in $K''$ with a sufficiently high probability.

Let us simply consider a PAC from using the Leftover Hash Lemma (35) because it is the only known way of quantifying actual security levels. Even more generally, from the nature of privacy amplification as bit sequence compression, we can see that a long $K''$ prior to compression is needed to obtain a good security level using a sufficiently high compression bit ratio. In contrast to ECC, one can break up $K'$ into shorter pieces and compress each piece. There is no correctness constraint; the only fundamental limit is whether the $p_1(K)$ of the smaller pieces is sufficiently small to ensure security from (35), assuming that (38) holds. As discussed in Sections VI.A-VI.B, this assumption does not hold when the ECC is covered by an imperfect key, as in the QKD case. Apart from this crucial issue, (35) provides the fundamental exchange between the key rate and the security level, other than the need to use the Markov inequality multiple times, as discussed in Section V.A.    

The key point of this connection is that it is not possible to have a key that can be made arbitrarily close to perfect by a security parameter, namely, condition (ii) of unconditional security in Section II, from only the asymptotic vanishing of mutual information or statistical distance, as explained in Sections III.A and IV.A. With (35), the limit on the exchange is explicit. This limit can be relaxed in one direction by sacrificing security to obtain a better key rate with the $\epsilon$-smooth entropy formulation commented on in [43]. However, relaxing security is not satisfactory given the current inadequate values to be discussed in Section VIII. Furthermore, relaxing security for longer keys is what conventional cryptography is apt to do.                              

In practice, the use of long ECCs and PACs is limited by the complexity of the processing involved. Both ECC processing and large matrix multiplication have been studied for decades, and it appears that it will be impossible in practice to address ECCs on $K''$ broken into pieces, whereby each of which is significantly longer than $10^6$, in the foreseeable future. In the absence of a full protocol including message authentication, a QKD \textit{round} may thus be defined by the stages of the protocol that check QBER and generate a sifted key $K''$ with subsequent ECC and PAC applied, as in Fig. 2, the length $|K''|$ being limited by current technology.

A more important concept of a QKD \textit{block} $K_b$ may be defined by the PAC through (35), with security level $d(K_b)$ determined from $p_1(K_b')$ for input blocks of length $|K_b'|$ to the PAC. Thus, $|K_b'|$ has a maximum value of $|K'|$ for a round but may be considerably shorter. Under the assumption of $K'$ correctness, the discussion on $K'$ being broken down into many $K_b'$ is based on practical considerations. However, this impacts on security because the blocks $K_b$ within a round may be correlated, and we also need to bound $p_1(K_b')$ instead of $p_1(K')$.

\section{Limit on Use of QKD-Generated Keys in Message Authentication}

Message authentication, in which a data message is checked to determine whether it has been altered, is often considered as a cryptographic task more important than privacy [3], [4]. A QKD protocol necessarily involves message authentication on the open exchange between the users for basis matching, QBER checking, and concerning the choice of ECC and PAC after Eve sets her probe. At the very least, message authentication is required to thwart a man-in-the-middle attack Eve may launch by pretending to be Alice to Bob and Bob to Alice. The security of a message authentication code (MAC), which is a hash function for bit compression, is sometimes based purely on complexity. In QKD protocols, MAC has to have information-theoretic security; otherwise, it would contradict the QKD claim of being information theoretically secure. A review of information theoretically secure message authentication can be found in [55, ch. 4] and [56]. A brief summary for our purpose is given as follows.

For a data message $m$ of a given bit length, a data tag of much shorter bit length $t=h(m)$ is obtained by applying a ``hash function" $h$ to $m$, say, by a compression matrix, as in a PAC, which is chosen from a given family of hash functions. In a substitution attack in the open tag case, given $h(m_1)=t_1$ and $m_2$, Eve finds $t_2$ with $h(m_2)=t_2$.   If the $h$ is chosen with a uniform secret key $K^h$, Eve's success probability $p_s$ is bounded by $\epsilon$ when the family of the hash function is an $\epsilon$-ASU family. Concerning both  substitution and  impersonation attacks, in the latter, Eve finds $t$ for a given $m$ such that $t=h(m)$ for the correct $h$ with success probability $p_I$:
\begin{equation}
p_s \leq \epsilon \text{, } \text{ } \text{ } p_I \leq \epsilon
\end{equation}
There is a general lower bound on the achievable $\epsilon$ for a given tap bit length $|t|$ that may be achieved:
\begin{equation}
\epsilon \geq 1/|t|
\end{equation}

When the key $K^h$ is a QKD key with $d(K^h) \leq \epsilon'$, it can be shown that [57]
\begin{equation}
p_s \leq \epsilon + \epsilon'\cdot|t|
\end{equation}
which may go to 1 and be achieved with equality for some $t$. The average of $p_s$ over possible $t$s, $\overline{p_s}$, is bounded by $\epsilon + \epsilon'$, as is $\overline{p_I}$ [57]:
\begin{equation}
\overline{p_s} \leq \epsilon + \epsilon' \text{, }\text{ } \text{ } \overline{p_I} \leq \epsilon + \epsilon'
\end{equation}
It follows from (47) that $\overline{p_s}$ and $\overline{p_I}$, not to say $p_s$ for individual $t$s, cannot be decreased with longer $|t|$ or longer $|K^h|$ so long as the level of $d(K^h) = \epsilon'$ is given. In particular, the authentication security parameter $|t|$, which allows security to be arbitrarily close to perfect from (44), is \textit{lost} due to the use of an imperfect $K^h$. 

We do not yet know how to rigorously remove the average $d$ and average $t$ guarantee in (47) via the Markov inequality to obtain an individual guarantee because the problem is nonlinear. If we apply (33) as a guess, numerically, after averaging is removed via (24) for obtaining an individual $d(K^h)$ and an individual $t$ guarantee, one would need $d\sim 10^{-30}$ to achieve the same security as a 32-bit $|t|$ from (45) and (47) or $d \sim 10^{-60}$ on a 64-bit $|t|$.  These $d$ values are completely unrealistic, as observed in Section VIII. Note, however, that that is simply the $p_1(K)$ level and does not cover BER leaks that provide information  on $K^h$ to Eve. 

Similar results are available for multiple uses of a hash function with tags covered by an imperfect key $K^t$ with $d(K^t) = \epsilon''$, say, for $m$ uses of $h$ [58]:
\begin{equation}
\overline{p_s} \leq \epsilon + m\epsilon''
\end{equation} 
A large number of uses of $h$ are needed in one QKD round because many uses are needed for authenticating a long sequence of bits. Thus, the security guarantee is further lowered with (48). Its quantitative effect on the final security of $K$ is unknown because the universal composition argument is not valid due to the problem nonlinearity alone. 

Equally significantly, the authentication steps in a QKD protocol have not been specified within the context of the other steps, with its imperfect levels considered. As we have seen, one cannot obtain a valid derivation of the final $K$ security level by declaring universal composition without explicitly detailing the justification of the argument in context. The execution of a QKD round requires a significant number of shared secret key bits for message authentication and error correction. It is yet unclear how quantitative security would emerge if the key used for such purposes is imperfect, as it must be for most QKD rounds.

​\section{Numerical   Inadequacy   Of   Security Guarantee}   ​

As emphasized in Section IV.A, the criterion $d$ applies to a key $K$ generated in a single QKD round. It would \textit{not} be meaningful to cite a $d$ level without stating the length of $K$ to which it applies, in effect making it a $|K|$-dependent $d(K)$. Eve's maximum possible probability of obtaining the entire $n$-bit $K$ is, from (14),
\begin{equation}
p_1(K) = 2^{-n}+d
\end{equation}

\subsection{QKD Block versus QKD Segment}

For security quantification, it is important to make the following two distinctions. First, as discussed in Section VI.D, the key length $|K|$ in $d(K)$ may refer to that of a QKD round or to that of a QKD block. Second, within a block, we can have many different mutually disjoint segments $K^*$ of consecutive bits under attack by Eve with bit gaps between them. Security is prescribed by $d(K_b)$ whether $K_b=K$ or not. The maximum probability of leaking $m$ different $K_i^*$, $i = 1,..,m$, within a block is given, from (13), by

\begin{equation}
p_1(K_1^*,...,K_m^*) = 2^{-\sum\limits_{i}|K_i^*|}+d(K_b)
\end{equation} With KPA conditioning, Eq. (50) is replaced by an average over $K_1$ on the left-hand side similar to (23) with respect to (13).

Note the nature of Eq. (50) in contrast to the failure probability per bit $d(K)/|K|$ of [35],[36], with such bit failure probability taken to be independent among bits of a block $K_b=K$, as in (20). The latter vastly underestimates the $p_1(K_1^*,...,K_m^*)$ of (50). In particular, it is not the case that a single bit would be leaked with probability $d(K_b)/|K_b|$, as the failure probability per bit interpretation implies. Rather, a number of bits equal to the block length $|K_b|$ is leaked with probability $d(K_b)$ (apart from the $2^{-|K_b|}$ factor).

Generally, it is misleading to evaluate key rate or security level  on a per-bit basis. The actual data rate per unit time should be employed for practical assessment, as is the compromise probability per unit time. To bound Eve's success probabilities, one may not assume that the blocks $K_b$ within a round are independent; however, one may make such an assumption for the $K's$ from different rounds. Certainly, the segments $K_i^*$ within a round are not independent, as shown by (50). Even when assuming that the blocks $K_b$ are independent, the segment's total compromise probability (50) is far larger than that given by the failure probability per bit interpretation. The division of $d(K_b)$ by $|K_b|$ is fortuitous and misleading.

The numerical solution is illustrated in the next subsection.

\subsection{Numerical Values}

We consider here the use of many keys $K$ generated in different protocol rounds of a QKD system to guarantee that the worst-case parameter $p_1(K)$ from $p(K)$ from each round is used, considering that there is no average on $p_1(K)$ itself, because there is no distribution for the complete $p(k)$. (See also objection B in Appendix III.) When there is a distribution on a random system parameter, we do not employ an average as a security measure for reasons discussed in Section V.C. (See also objection C in Appendix III.) In particular, according to the operational guarantee statement (OG) of Section III.A, a finite sample average in an experiment is a random quantity on which probability statements can be made instead of approximating it using the nonrandom average.

Although the maximum $p_1$ of $p(k)$ is an unknown nonrandom parameter without a distribution rather than a random parameter [39], it is the probability of an event, and we can talk about averages or expectation values [41]. We cannot estimate its spread as in the random parameter case and will simply consider the probability as a fractional average. This is in contrast to the situation wherein a distribution exists and a Markov inequality (29) can be used to produce an operational probability statement from the average (mean). In the present case, the collection of $p_1$ in different rounds can be given instead. The following average values cited are to be understood as having the same import as a probability strictly speaking. Here, we cannot disentangle the possible conceptual subtleties of probabilities in real-world applications. (However, see [41].)

The theoretical numerical values of [29] for single-photon BB84 provide a tradeoff between key rate and security.  $d(K_b) \sim 10^{-9}$ is obtained for $|K_b| \sim 10^5$. If the key rate is $10^7$ bps, a segment leak of a block or a total of $10^5$ bits within a block may be leaked on average every 100 days if the $d$ level is individual. The leak becomes 300 blocks per day after one use of the Markov inequality from (30). Against a KPA, the average leakage becomes one block every 10 seconds from (33).

The experimental results in [59] give a key rate of $1.4\times10^5$ bps and $d(K_b)\sim 4\times10^{-9}$ for $|K_b|\sim10^5$, which are approximately equal to the values given in [36]. This amounts to an average maximum of 6 blocks or $6\times10^5$ bits per day leaked against ciphertext-only attacks. Against a KPA, the rate is 100 blocks or $10^7$ bits per day.

Note that, as shown in (50), these leak levels could apply to many different segment leak combinations distributed across a single block. If the segments spread across more than one block, one should compute the leak probability of those within a block from $d(K_b)$ and then multiply them from independence to obtain the total joint leak probability. Only observing the block-leak probability, it appears that the above numerical guarantee is far from adequate for almost all applications and is certainly not adequate for commercial banking.

It is often argued that $10^{-15}$ is the ``practical" probability level for guaranteeing impossibility. With realistic numerical values of $10^7$ blocks per day with $|K_b|=10^5$, a $d$-level guarantee of a practically perfect key (but only from the viewpoint of $p_1(K_b)$, i.e., not ``failure probability") in one day of operation would require a $d$-level of $10^{-44}$ for  ciphertext-only attacks alone. Such a level is 35 orders of magnitude above the available values of $d$ from theory alone. There is no indication of how the numerical gap can be closed in any significant manner, again only in theory and even assuming that the  security analysis in the literature is completely valid, which is not the case. On the other hand, security can be increased to a near-uniform level by further sacrificing key rate, as presented in [9], although that would require a larger $|K|$ than the literature value, which is limited by ECCs and other processing complexities; see Section VI.C.

According to the failure probability per bit interpretation discussed in Section IV.A under statement (F), a failure probability per bit $d/|K|$ of $10^{-24}$ for all generated bits  means that the QKD protocol ``can be run for the age of the universe and still have an accumulated failure strictly less than 1". This conclusion is obtained from the incorrect $P_f$ of (20). The block value $|K_b|$ is not specified. We take $|K_b| = |K| = 10^6$ for 1 second of operation, which is a sensible value. These numbers imply a possible average leak of $10^4$ bits per day for ciphertext-only attacks and 100 bps for KPAs. This strongly contradicts the quote (F) that the accumulated ``failure" is strictly less than 1 over the age of the universe.

\begin{table}[!t]
\centering
\includegraphics[width=3in]{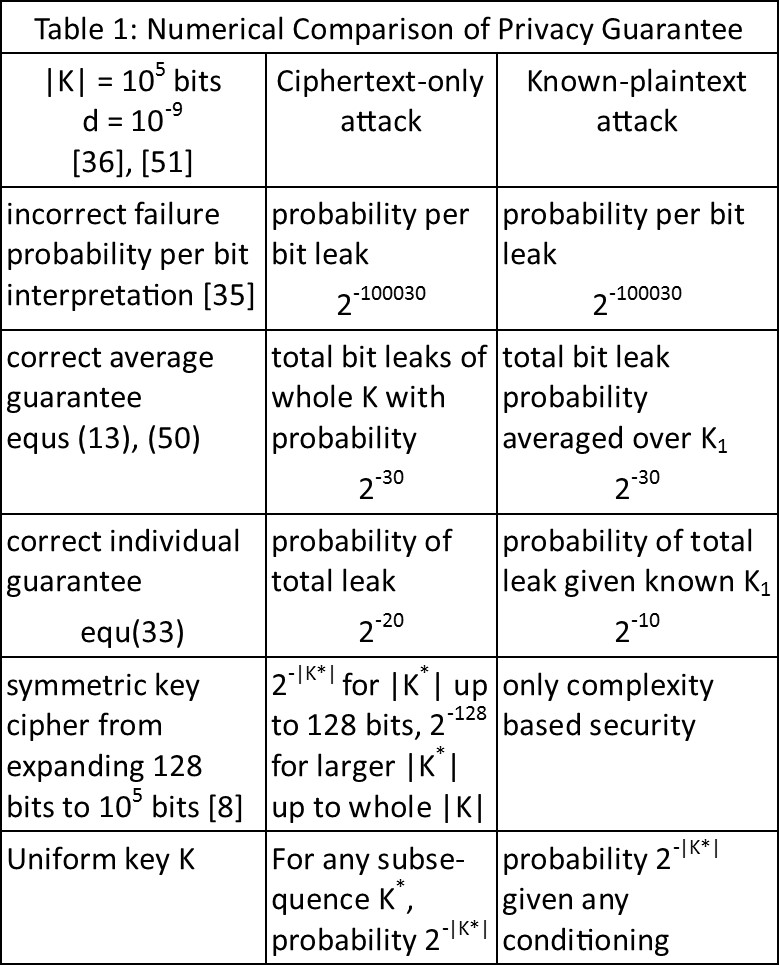}
\end{table}

Table 1 compares the numerical guarantees for a $|K_b| = |K| = 10^5$-bit block at $d=10^{-9}$ and the current theoretical [36] as well as experimental [59] values at Mbps key rates. There are five cases that can be compared: the incorrect failure probability per bit interpretation [35], the correct average guarantee from Eqs. (13) and (50), the individual operational guarantee Eqs. (30) and (33), the case where $K$ is uniform, and the symmetric cipher results (see App. II) from expanding a $128$-bit seed key to $10^5$ bits. For this nearly $1,000$-fold symmetric key expansion, the seed key cost of $\sim128$ bits per $10^5$ bits is needed in QKD for message authentication within the QKD key generation round of $K$. These results show that, even simply as an average guarantee, there are only $30$ bits of security in the QKD system for the $10^5$ bits compared to the $128$ bits of security for conventional ciphers under ciphertext-only attacks. Further discussion is given in Appendix II. The relevance of known-plaintext attacks is discussed in Appendices II.B and II.C. 

\section{Summary and Conclusion}

In this paper, we do not address the problems of the very low efficiency of QKD, some of which are discussed in [60] and many of which are intrinsic due to the small signal level and resulting sensitivity to disturbance. More significantly, we do not address the physics-based security issues that have not been addressed in the QKD literature but which are fundamental to a valid security claim [61]. In addition, especially important are the detector hacking attacks [62] that break most current QKD implementations, including the checking of the Bell inequality in establishing EPR pairs [63]. This demonstrates that cryptosystem representation is a tricky and difficult issue in physics-based cryptography, especially in QKD, where many physical details may affect a single photon or a small signal that would not matter for stronger signals. It is not yet clear what would be a reliable justification to ensure that a particular QKD system representation has incorporated all the essential features of simply the cryptosystem operation in a mathematical model that would address hacking, in addition to other extraneous loopholes. This is a point first emphasized in [64] and is found to be prescient in several ways.

In this paper, we have analyzed the fundamental information-theoretic security guarantees in cryptosystems and shown in what ways current QKD security analysis falls short. A brief history of some security works in the literature is given in Appendix I, which also contains a summary of the contents of the different sections in the body of the paper. In Appendix II, a brief comparison of QKD with conventional cryptography is given to put the significance of QKD into perspective. In Appendix III, some possible points of objection or confusion are addressed.

A most important point of our foundational analysis is that security must involve Eve's success probabilities in various problems. The incorrect failure probability interpretation dissected in Section IV implicitly recognizes such importance, and it has the following faulty security consequences:

\begin{enumerate}[label=(\roman*)]
\item Known-plaintext attack security is incorrectly quantified, as shown in Section V.A.
\item The important criterion concerning Eve's bit error rate is incorrectly bounded, as described in Section V.B.
\item Universal composition is obtained when it is not valid, as in known-plaintext attacks, or when it requires a further justification that appears impossible to provide due to nonlinearity, as in error correction treated in Section VI.B.
\item The security situation in message authentication is misrepresented as an individual guarantee, as discussed in Section VII.
\item The failure probability per bit interpretation is seriously incorrect, as discussed in Section IV.A with numerical security levels illustrated in Section III.B.
\end{enumerate}
Generally, the failure probability interpretation ascribes substantially improved quantitative security to what can be validly deduced both qualitatively for problems in which the trace distance criterion is yet to provide a guarantee and quantitatively to problems the interpretation does give a guarantee to by neglecting the difference between average and individual guarantees.

It appears that current QKD security is fundamentally no different than the uncertain security of conventional mathematics-based cryptography. One may offer plausibility arguments for security and quantify security under some restrictive assumptions; however, there is no proof against all possible attacks. It may be useful to conduct research to develop new features for a QKD system that would permit a general security proof that is both transparent and valid. It would also be useful to utilize quantum effects on larger signals to obtain information-theoretic security. Some such attempts have been undertaken in [5] and [65] in the KCQ and DBM approaches. It remains to be observed the extent to which QKD can be so broadened usefully.

\appendices
\renewcommand{\theequation}{\thesection.\arabic{equation}}
\newpage
\section{History of QKD Security Proofs}

I would like to begin this appendix with the following quotations:\\
``The variety in this field is what makes cryptography such a fascinating area to work on. It is really a mixture of widely different fields. There is always something new to learn, and new ideas come from all directions. It is impossible to understand it all. There is nobody in the world who knows everything about cryptography. There isn't even anybody who knows most of it. We certainly don't know everything there is to know about the subject of this book. So here is your first lesson in cryptography: keep a critical mind. Don't blindly trust anything, even if it is in print. You'll soon see that this critical mind is an essential ingredient of what we call ``professional paranoia." [4, p. 3]\\``it is very easy for people to take criticism of their work as a personal attack, with all the resulting problems." [4, p. 10]

These words were written on conventional cryptography. They are even more appropriate for QKD. 

In this appendix, we briefly outline the history of security proofs on BB84-type QKD protocols. There are a very large number of papers on security proofs in QKD, many of which are referenced in [2]. We will touch upon mainly those that have been mentioned in the body of the paper, including the more influential proofs on the security quantification of concrete QKD systems. We will also take the opportunity to mention some relations between security analysis and QKD experiments thus far and to discuss some major physics security issues not addressed in the body of the paper. Security proofs for BB84 are the most well developed in the field. Other security proofs share almost all the difficulties BB84 proofs face and more. We will summarize at the end a list of problems that no proof in QKD has yet overcome​, with the exception ​of the KCQ-DBM approach; however, the details of why and how that is possible are yet to appear.          
        
It may be noted that security proofs, in QKD or any cryptosystem concerning privacy and key distribution, are a very complicated matter. Errors and incompleteness are to be expected during the early stages of their development. These theoretical defects cannot be glossed over in cryptography, although ​such defects​ ​are​ often justifiably ​neglected​ in physics and engineering when a final working experimental system is what decides success or failure. Security cannot be proved experimentally, if only because there are an infinite variety of possible attacks​, which cannot all be described​. There were many surprises in the history of cryptography; thus,​ whether there is a valid proof in an important issue, especially in QKD, where provable security appears to be the only real advantage compared to conventional cryptography.  

As in the case of many mathematical propositions, it is not always possible to produce counter-examples to the main conclusion. Sometimes, the statement is actually true, such as the Poincare Conjecture and Fermat's Last Theorem, yet a valid proof is a separate matter from assuming the truth. In the body of this paper, we could only produce counter-examples to specific spots of reasoning in a purported proof. We did not give a specific attack that would always succeed. The burden is on those who claim that there is a proof to produce a valid one. One can always change the proof claim to a plausibility claim, and we need to draw sharp boundaries in cryptography. The discussions of this appendix should be read with th​is​ in mind.

\subsection{Earlier Proofs}

The earliest general BB84 security proofs in [13] and [26] are mainly on the security of the sifted key, namely, $K''$ in Fig. 2. Earlier versions of [13] appeared a few years before it did, and [66] provided an important direction for [26]. There are several noteworthy problems in these proofs, some of which are misinterpretations from others​ and​ not by the authors; however, ​such errors ​have perpetuated.
 
\textit{First}, these proofs are asymptotic existence proofs asserting the existence of a protocol that would yield a purportedly perfect key in the limit of long bit length $|K''|$. They use the mutual information criterion, which we have shown ​in Section III.B ​cannot lead to such a conclusion ​by its mere vanishing asymptotically​. This conclusion​ could not ​be drawn  with the trace distance $d$ going to​ $0$ ​either (Section IV). However, the prevailing impression is that ​it​ could, and the issue is not addressed in the recent review [2].

\textit{Second}, there is no treatment of known-plaintext attacks when the generated key is used for encryption. Apparently, the quantum accessible mutual information criterion is not sufficient for proving security against KPAs, as discussed in Section IV. A weakened protection against KPAs is provided by $d$, as presented in Section V.B.

\textit{Third}, Eve's side information from error correction and privacy amplification are not considered and were later addressed in different approaches, as discussed in Section VI. The ECC problem remains to be rigorously treated ​for​ any type of ​QKD protocol, of which we call QKD or otherwise. 

\textit{Fourth}, these proofs are on qubits (two-dimensional quantum state spaces) and lossless systems. In all implementations, we have infinite-dimensional photon state spaces with loss. For example, coherent detection by Eve is ruled out by the qubit model. Loss is ubiquitous in optical systems. N​o​ reason has been offered as to why it would only affect throughput but not security in BB84, although it is known that it does affect security in B92​ [1]​. See [61] for further discussions of these and other physics-related security issues.

\textit{Fifth}, although [26] is an existence proof among the class of what is called CSS ECCs with associated privacy amplification, it has been widely taken to have proved the (asymptotic) security of any specific ECC and any PAC. This error is found later in both experimental and theoretical studies.

There are various spots of uncertain validity in the reasoning of these papers. Although they are relevant to security, for the sake of this paper, we can assume that they are valid. The main concern in this regard is that the issues involved are not purely mathematical but concern the relation of a mathematical statement to its real-world implication. We have observed some such examples in Section IV for cryptographic relations. It is a special problem for QKD in which quantum physics at the small scale is tied to various classical physical or engineering phenomena.

An important sequel to [26] is the widely quoted [27], which extends [26] to include various system imperfections by adjusting the final result in [20] using the attainable key rate with purported asymptotic perfect key generation. The derivations of these adjustments are brief and heuristic and are based on ad hoc estimates. There is no general formulation of the problem including an imperfect feature that would demonstrate how the original proof would address all possible attacks with such imperfection. The PAC in [26] is a nonlinear hash function; however, is treated as if it is linear. This [27] is used as the basis of the security claim on the use of decoy states for laser instead of single-photon sources; some problems with such a connection are discussed in [61]. In particular, it is not realized that a weak laser pulse is itself coherent and not a mere multi-photon qubit [61]. Ref [27] is also used in the security claim of the so-called measurement-device-independent approach [25].

\subsection{Later Finite Protocol Proofs}

Security proofs for a finite and more specific protocol were developed and culminated in [36] for lossless BB84 with various imperfections. Many approaches to bounding Eve's information on the sifted key $K''$ have been attempted, therein settling on the ``smoothed" minimum entropy, which is used in numerical evaluations in [36]. Such smooth entropy is equivalent​ly​ Eve's maximum probability of obtaining $K''$ ​but ​with greater flexibility in terms of giving up some level of security for a higher key rate. ​(The use of these smooth entropies cannot increase security by lowering the key rate.) ​The trace distance criterion $d$ is used​ because the small KPA leak in the example of [31] was already considered unacceptable, and the incorrect failure probability claim from a $d$ guarantee was maintained. We have discussed in Sections V.D that $d$ and accessible information are indeed very different guarantees in the quantum domain but are essentially equivalent classically from (34).

The errors in misinterpreting $d$ are analyzed in Sections IV-V. Quantitatively, the numerical values of $d$ that were obtained ​are​ far from adequate simply ​on​ the probability of compromising the entire generated key $K$ in a block, as discussed in Section VIII. The actual security guarantee from $d$ is ​detailed in Section V. It is ​not given by the incorrect failure probability interpretation​, and it is not known whether it can cover BER leaks, which for example Eve could use to attack the QKD-key-covered parity check digits of a linear ECC discussed in Section VI.B​. The PAC information leak is fully considered in [36], although the ECC leak is not, as discussed in Section ​VI. There is the  serious problem of using an imperfect key for the purpose of covering the ECC parity check digits​  mentioned above​ and for message authentication in future rounds, ​the latter​ being discussed in Section VII.

The approach of [26] is generalized in [45], [51], [52] for a finite protocol. It is difficult to assess the formal results​ in these papers​.​ In [52], the same ad hoc formula for ECC information leaks is used in an actual evaluation, as in [36].​ In a concrete protocol, there is are advantages,  only disadvantages, to​ these CSS-code-based​ approaches in addressing the ECC and PAC information leak problems ​as ​compared to the approach of [36].

\subsection{Relations​  to  Experiments​}
​ ​
We will simply provide some general remarks to indicate certain problems in the QKD experimental literature and will not dwell on specific analysis of the errors in specific papers. That is a separate subject matter, i.e., not that of basic security analysis. Given the complexity of QKD security analysis, it is a highly nontrivial task to integrate all the components of a protocol for an experimental system.

To begin with, no complete QKD protocol that includes message authentication and error correction​ with an imperfect QKD key has been analyzed. In particular, the message authentication steps are not interlaced with exactly how the bulk of the protocol runs or with what is being authenticated at what time, and the ECC information leak is only considered  by an ad hoc formula without considering the relatively large $d$ level of the key used to cover its parity check digits. QKD experiments do not usually concern the entire cryptosystem, the necessary message authentication, or error correction and privacy amplification.  Often, a key rate is cited with no security level attached, which is nonsensical for a concrete protocol, as we observed in Sections II-III. Part of the cause is apparently the use of [26], as discussed in the first point of Appendix I.A, with the belief that security can be made arbitrarily close to perfect for a given key rate below the threshold formula of [26] or [27].

Such key rate results from [26], [27], with or without system imperfections and assuming that they are completely valid, have yet to consider ECC and PAC leaks. More significantly, they are often quoted for a system that employs error correction and privacy amplification methods, which is not a CSS code. Thus, those formulas so quoted are not relevant because they have never been shown in any way to hold outside of CSS codes, and even then, such proofs are merely existence proofs and do not pin down the working codes.

The situation is evidently better for the approach of [30], the problems of which we have analyzed in the bulk of this paper and briefly mentioned in Appendix I.B. Even if we assume that everything is valid, the obtainable security level is insufficient, i.e., $d$ is too large for many applications, as discussed in Section VIII. Although a $d$ level guarantee is not complete, it appears to be a useful criterion and needs to be ascertained for any concrete protocol analyzed without giving Eve's full success probability profile (4). However, DBM [65] promises a new direct security approach yet to be made public.

\subsection{List  of  Major  Unsolved  Problems}

There are three major security analysis problems that have not yet been solved for any QKD protocol, with exceptions noted below.

\begin{enumerate}[label={(\arabic*)}]

\item In the presence of inevitable losses, it has not been proved that only throughput is affected but not security.

\item When using an imperfect key in executing a QKD protocol, it is not known what the error correction information leak would be.

\item No analysis has ever been given on a full protocol involving message authentication with an imperfect key, therein demonstrating the effects of key imperfection on the security level of the final generated key.

\end{enumerate}

Note that point (1) does not apply to CV-QKD (continuous variable), which is also immune to detector blinding attacks [62]. (We do not discuss such very serious hacking problems in this paper.) However, CV-QKD suffers from other major problems [61] not found in BB84.

The recent approach [67], which allegedly dispenses with the information-disturbance tradeoff without considering even intercept-resend attacks, is subject to all three points. The coherent-state KCQ approach in [5] is not subject to points (1)-(2) and may not require error correction due to its substantially larger signal level. With error correction, the new DBM technique [56] may be required.

\section{Comparison  Of  QKD  And  Conventional Cipher  Security}     

By ``conventional ciphers", we mean mathematics-based ciphers, which cover essentially all practical ciphers in commercial use [3], [4]. These are different from ``classical ciphers", which rely on simply classical physics and include physical-noise-based cryptography such as that described in [18], [19] and [20]. Quantum-physics-based cryptography is yet another gene; however, by QKD, we mean the smaller subset defined in Section I with BB84 as representative. QKD covers key generation and direct encryption with the generated key. We will compare both to some typical conventional ciphers in current use. Such a comparison is important for assessing the potential, progress, and  future of QKD.

\subsection{Asymmetric  Key  Ciphers}

QKD has often been contrasted with asymmetric or public key cryptography, which only includes complexity-based security and no information-theoretic security other than that in the sense of (7) in Section III.A. However, it is substantially more appropriate to compare QKD with symmetric key ciphers [8] because a pre-shared secret key is needed to execute a QKD protocol other than for the purpose of agent identification. Message authentication is needed to prevent man-in-the-middle attacks. For this purpose alone, one would need information theoretically secure message authentication to preserve the overall information-theoretic security of the QKD protocol, which requires a shared secret key. As shown in Section VI, the error correction step of a QKD protocol also requires a pre-shared secret key. Of course, the QKD-generated key represents a pre-shared key for future protocols. We have shown in Section VI how the imperfect security level of such a key prevents a valid security proof and its quantitative level from being obtained.   

In this subsection, we will simply make some comments on the contrast between public key cryptography and QKD. Asymmetric key protocols can be used for both key generation and direct encryption for privacy. They are not used for the latter in practice due to their relative inefficiency compared to symmetric key ciphers. QKD in practice is more inefficient and more complex to operate compared to asymmetric key ciphers due to a number of fundamental reasons such as low signal levels and inevitable large losses. They have the advantage of being provably information theoretically secure, which is however not yet realized, as we show in this paper. 

The advantage is often claimed that QKD encryption is resistant to future compromise of the secret key in conventional ciphers. That is surely the case in comparison to asymmetric key ciphers because decryption of the public ciphertext may be obtained in the future based on mere computational power. However, QKD has no real advantage in this regard compared to symmetric key ciphers because the shared secret key can simply be  deleted permanently. Furthermore, when the generated key from QKD is used on a conventional cipher, such a key shares the same problem, if any, as in the case of symmetric key ciphers.

\subsection{Symmetric  Key  Ciphers}

Symmetric key ciphers can be used for ``key expansion", effectively generating new ``session keys" from a master key, or for privacy encryption. When used for key expansion, they are very similar to QKD generation schematically. They have information-theoretic security because they are similar in their security as the plaintext security under ciphertext-only attacks when the cipher is used for encryption. Specifically, the generated (running) key sequence $K_r$ from the cipher with a uniform seed key $K_s$, which can be regarded as a pseudo random number sequence, would have the following probability of leaking the whole $K_r$ to Eve:

\begin{equation}
P(k_r) = 2^{-|K_s|}
\end{equation}                                                 
This is obtained because each possible $K_s$ value leads to a different $K_r$ sequence (non-degenerate cipher), which may be used in Eq. (1) as the $K$. It is important to realize that this \textit{is} information-theoretic security for $K_r$, and it is very favorable for typical values of $|K_s|$ from $100$ to $1,000$, as compared to that obtained from the QKD value (50) with the $d$ values in the literature. For comparison to QKD, a block cipher can be run in stream cipher mode for the generation of a running key, as in Fig. 1.

Subset leaks $p(k_r^*)$ depend on the specific conventional cipher. For (non-degenerate) linear feedback shift registers [3], the level is perfect for a single $K_r^*$ sequence of $\leq |K_s|$ consecutive bits [6]. In general, the correlations between bits in $K_r$ are difficult to quantify, whereas a QKD key obtains a security guarantee under (13). In any case, key expansion symmetric key ciphers do not have ``perfect forward secrecy"​​[3] due to the correlations between bits in $K_r$. Moreover, a QKD-generated key does \textit{not} have such secrecy either, especially not at the large $d(K)$ level given in the literature,  because it is imperfect.

The following numerical comparison of the QKD system of [59] with only a linear feedback shift register (LFSR) cipher against ciphertext-only attacks is revealing. With a seed key of only 128 bits, the $p_1(K_r)$ level of an LFSR is $\sim 10^{-40}$ from (II.51) for any $|K_r|$, which compares quite favorably to $\sim 10^{-9}$ for $|K|\sim10^5$ bits in [59] even before the Markov inequality is applied for an individual guarantee. The LSFR protects a segment $K^*$ of up to 128 consecutive bits with perfect security, whereas the system of [59] only does so at the same $10^{-9}$ level from (13). It is not known what the LSFR information-theoretic security is for many scattered segments $K^*$, and [59] gives the same $10^{-9}$ probability for segments within a block from (50). There are many other uncertain securities in both systems. It is also not clear if one is superior to another security wise. However, it is clear that the LFSR is substantially faster and cheaper to operate. The numerical comparison of QKD and symmetric-key ciphers is included in Table 1 of Section VIII.B.

Note that  there is no KPA in key expansion until the expanded key is used in an application because the plaintext is chosen to be $U$, at least in principle, by the encrypter. This is why there is information-theoretic security in conventional key expansion before the key is used, given the possibility of KPA. Non-degenerate symmetric key ciphers in current use do not have any KPA information-theoretic security because only a length of $|K_s|$ known input bits together with the corresponding ciphertext would uniquely fix $K_s$ in principle. Security relies on the complexity of locating $K_s$. On the other hand, the QKD key remains secure for sufficiently small $d$ from (23). Note that it may be possible to obtain information-theoretic security against KPA with a conventional cipher. The theoretical possibility is presented in [68, App.], especially when the known plaintext is not too long.  

It has been proposed that the QKD generated key can be used as the seed key in a conventional cipher. In that case, the plaintext so encrypted only obtains the protection of the conventional cipher but worse considering that $K_s$ is no longer perfect. How the imperfection affects the conventional symmetric key cipher security is \textit{unknown}. In any case, as a pure conventional cipher, there is no more information-theoretic security against KPA. 

\subsection{Relevance of KPA and Kirchhoff's  Principle}

We believe the following remarks are important when comparing QKD with conventional cipher security. In many specialized applications, it does \textit{not} seem possible to launch a KPA, in contrast to most commercial applications. Examples include military applications with encryption on board an aircraft, a ship, a satellite, or a protected ground station. In such cases, it is not clear what advantage of significance QKD provides compared to conventional encryption, as discussed in the above subsection with a numerical example. This is especially true given that the QKD security advantage has yet to be rigorously established; in addition, it is inefficient and is vulnerable to hacking. 

More broadly, for such specialized applications, it is not clear why Kirchhoff's principle [​4​] should be assumed. That principle states that the o​nly security​-relevant​ feature of the cryptosystem that an attacker does not know and that the users do is the shared secret key between the users. The cipher structure and the encryption algorithm are assumed to be openly known. This does \textit{not} appear to be a reasonable assumption in military situations. If the encryption structure​ or​ algorithm is unknown to the attacker, it appears next to impossible for her to obtain substantial amounts of information for any reasonable cipher the users choose  because the possibilities ​between structures and algorithms a​re endless​ and​​ equivalent to a huge number of shared secret bits.​​ They can be readily and often changed under software implementations.

Even under KPA and Kirchoff's principle, there is no known vulnerability of conventional strong ciphers such as AES. In specialized applications, a huge number of seed key bits can be pre-stored. Weaker ciphers are commonly employed due to their efficiency. The notable security risks are not from the known strong ciphers. Is there a serious problem that awaits QKD as its solution? It appears that efficient bulk encryption of large (elephant) data flows in optical links is the one clear area that would benefit from efficient QKD.

\section{Objections and Answers}

This appendix addresses some possible objections or concerns on various points of this paper.
\\

Objection A:  Security is a matter of definition. Why is your definition better than other ones?

Answer:  Security is not a matter of definition. The cryptosystem designer must decide on an acceptable probability of a successful attack by Eve on any characteristic of the generated key $K$ from a QKD round. Consider for example $K$ of length $10^5$ bits. If Eve has a total compromise probability $p_1(K) ~ 10^{-10}$ (for $d=10^{-10}$) of correctly identifying the entire $K$, which is substantially larger than the uniform level of $~10^{-30,000}$, is this acceptable? Suppose that it is not; then, regardless of the security definition used at any quantitative level, security is not guaranteed if the total compromise probability above is not ruled out. This is formalized by the operational guarantee statement (OG) in Section III.A. A theoretic security criterion has to yield operational probability guarantees, which must be the concern of cryptographic security. Such an operational guarantee is difficult to obtain and has been ignored in QKD, except through the incorrect failure probability interpretation discussed in detail in Section IV. It is also neglected in some but not all information-theoretic security studies in conventional cryptography.

In this paper, we detailed some basic operational guarantees for the trace distance (statistical distance) criterion $d$; however, not all important operational guarantees have been covered by $d$. In particular, Eve's BER, discussed in Section V.B, is not covered. When Eve identifies $K$ incorrectly as a sequence, she may still correctly obtain, say, 60\% of the bits, similar to the case whereby the distribution $p(K)$ is known to Eve with a per bit error probability of 0.6, which would not be considered a secure key by most designers. Most designers would want a proof against such a possibility at any designed BER level.

There are questions concerning the average versus worst-case guarantee, average versus individual guarantee, and security of multiple uses of different keys at given d levels. These questions are discussed in Section III.A, V.C, and VIII.B as well as in the following objections B and C. Note that Eq. (50) with $d=10^{-10}$ implies that many bits and bit segments may be leaked for operation in one, say, at a key rate of 1 Mbps; see Section VIII. 
\\

Objection B:  The average instead of the worst case should be employed in quantifying security leaks.

Answer:  For a rigorous assessment of a problem on performance depending on a parameter $\Lambda$, usually, only a relevant upper or lower bound can be obtained over the range of values that $\Lambda$ may take. The worst-case performance, say, concerning the time complexity of an algorithm or the security level of a cryptosystem provides a guaranteed level of performance that may or may not result from an attack but that cannot be exceeded. If the parameter $\Lambda$ has a probability distribution, one can also discuss the average performance; see Objection C.

However, it may not be meaningful, in the sense of being applicable to reality, to assign a probability distribution to what is called a ``\textit{nonrandom parameter}" [39], which is not described by a probability distribution. This may occur when the parameter appears in only one sample instance with no repeated trials (although there remains meaning to assign probability to such a situation in various theories of probability [41], such as the probability that President John Kennedy was shot by more than one gunman. The Warren commission addressed such question. See also Objection C.) In decision theory, an unknown nonrandom parameter ​[39] ​is ​then ​used. This ​often ​happens, for example, when the parameter takes on a continuum of possibility.

In this paper, the unknown parameter is $p(k)$, namely, Eve's distribution on the generated key $K$ from her attack, as presented in Section III.A as well as in the beginning of Section IV. The function is the parameter $\Lambda$ under consideration, the range of which has a cardinality of the continuum. More significantly,​ Eve can pick any attack for which there is no distribution, and in any event, the users do not know the distribution or if one exists. 

Thus, we cannot average over $p(k)$ or $p_1$, the maximum value of $p(k)$ from Eq. (4). We also cannot average over the $k$ of a specific $p(k)$ even though that may make sense if only because we do not know the value of that specific $p(k)$. Thus, we have to bound $p_1$ as the worst case to provide a valid guarantee.     
\\

Objection C: An average can be used for the guarantee instead of a probability. In particular, there is no need to apply a Markov inequality to convert an average guarantee into an individual guarantee.

Answer: Some parameters in QKD do have reasonable probability distributions, although only for a given attack in a given round. Thus, the choice of PAC is taken to be uniform. The known part ​$K_1$ ​to Eve of $K$ in a KPA is specified by the marginal distribution of $K$ from the joint distribution $p(K,Y)$ with Eve's observation $Y$. In Section V.C, we detailed th​e​ main reasons why a probabilistic guarantee is more accurate than an average guarantee. One reason is that the average has no operational meaning when the total number of trials (pertaining to that underlying distribution) is small, similar to the single-trial case. (Think of the above Kennedy assassination example​ in objection B.) This is codified in the statement (OG) in Section III.A on operational guarantees.

Equally significantly, a finite sample average remains a random quantity with its own probability distribution. A probability statement can be made on it to satisfy (OG). For example, an estimate from variance information could lead to such an estimate, not further information on the distribution. The Markov inequality estimates (30) and (33) from the average alone are weak because no other statistical information is available. 

One may be stuck with a weaker guarantee, such as the average without a probability statement, and even simply relying on a single theoretic criterion without analyzing its proper operational meaning, as has been the case in QKD until now, if that is all one can obtain. However, comparing quantitative security on various characteristics of $K$ to the uniform $U$ is the concern of rigorous security. A uniform $K=U$ gives far better and far more detailed security guarantees than does a trace distance guarantee, especially at the relatively ​large level that can be obtained. At the very least, it cannot be claimed that the QKD-generated key is ``perfect", can be made as close to perfect as desired, or is perfect except for a small probability. The many problems presented in this paper should make clear the dangers of such an exaggeration.
\\

Objection D:  Distinguishability advantage is a great criterion in cryptography. Is there ​a​ definite counter-example on why it is not satisfactory?

Answer:  Distinguishability advantage is a vague and very misleadingly phrased security guarantee. As detailed in Section IV.B, it leads to the an incorrect failure probability interpretation of a statistical distance guarantee (which the trace distance criterion reduces to upon Eve's measurement on her probe) as a definite ​and general ​quantitative consequence. Alone, it serves no purpose other than​ ​what is given ​mathematically​, namely, a bound on the statistical distance.

In particular, there can be no counter-example until one gives the quantitative guarantee that derives from distinguishability advantage. Using the failure probability interpretation as its consequence, all the counter-examples to the failure probability interpretation are counter-examples to the distinguishability advantage interpretation. These include the examples in Section IV.A and the KPA counter-example in Section V.A.

​Distinguishability advantage as a statistical distance bound on $\delta_E$ is a useful criterion, as demonstrated by ​this paper. It simply does not have the operational significance that has been ascribed to it. In particular, the $p_1(K^*)$ bound of Eq. (13) that results is the same for a one-bit subsequence $K^*$ of $K$ as it is for the whole $K=K^*$. This may give the impression that the cryptosystem is substantially more secure than it actually is and apparently led to the incorrect failure probability per bit interpretation discussed in Section IV.A, which grossly overestimates security.   
\\

Objection E: Isn't your KPA counter-example of section V.A not one of known-plaintext attack but one of chosen-plaintext attack?

Answer:  There is no difference between KPAs and chosen-plaintext attacks for the symmetric key additive stream ciphers of Fig. 1, and the counter-example concerns such a cipher. This is because the additive key stream in symmetric key ciphers is blind to the data. A KPA reveals part of the running key $k_1$ that happens to be uncovered from the known data $x_1$, with the following $k_2$ depending on $k_1$ and $p(k)$.

Of course, the $\delta_E$ level provides an average guarantee over $K_1$, as given in Eq. (23). Thus, a bad $k_1$ can only occur with a small probability for small $\delta_E$ (which $d$ reduces to). However, such an average needs to be removed for an individual guarantee. The incorrect failure probability interpretation produces an incorrect answer for a  given $k_1$; see Section V.A.
\\

Objection F:  The trace distance guarantee may be sufficient in practice. What is the evidence to the contrary?

Answer:  This paper is concerned with information-theoretic security foundation and rigorous proofs of security, the latter being proclaimed for QKD for almost twenty years. It is not clear what is meant by ``sufficient in practice", which would vary from application to application. Many problems, including the lack of a real proof on QKD security, are noted in this paper. They constitute evidence of possible practical security problems. There is no such thing as proof by ``no counter-example yet". The burden is on those who claim QKD has been proved secure to produce a valid proof for a given model.​ It is the task of cryptanalysis, a major component of cryptology, to scrutinize security, which is rarely performed in the QKD​ ​literature apart from implementation issues.
\\

​Objection G:  Why cannot security be brought arbitrarily close to perfect by privacy amplification?

Answer:  The trace distance level $d(K)$ is bounded by (36) in terms of the total compromise probability $p_1(K')$ of the shift key $K'$. It cannot be made arbitrarily small from the Leftover Hash Lemma. It is not known whether there is any way to make only $p_1(K)$ arbitrarily close to the uniform level $2^{-|K|}$; see Section VI.A. In addition, note from Sections III.B and  IV before IV.A that asymptotic vanishing of Eve's accessible information or the trace distance not only does \textit{not} imply that the key is arbitrarily close to perfect but also may even imply that the key suffers from a serious weakness of having a very relatively large $p_1(K)$.
\\

Objection H:  There is no problem in assigning a numerical value to $f$ in the error correction cost (39).
This can be taken from the actual ECC used in the pro​tocol​.

Answer:  ​In that case​, there is then no need to present  formula (39), which is irrelevant to the actual bit cost. This conveys the misleading impression that there is a general justification.

As outlined in Section VI.B, there is  a rationale for (39) when $f=1$ because of the asymptotic number of bits needed ​to cover a linear ECC ​for guaranteed error correction, alt​hough only for a binary symmetric channel, which is not obtained under a general attack. The point is that correctness of the round (the users agree on the same key) is then guaranteed. When a finite code is used with usually a bit cost of even less than (39) for $f=1$, correctness​ from an ECC​ cannot be theoretically guaranteed and must be established with high probability by other means not​ given ​in​ the security proof of the protocol. This is acceptable in practice whenever it works but cannot be confused with a security proof on the model. The assumption must at least be made clear that it is not logical​ly incorporated in​ the ​security ​proof.

The more serious problem with error correction is what is focused on in Section VI.B; it has not been quantitatively shown how an imperfect key covering the ECC would degrade security or why the imperfect key bit cost ​of any​ quantitative level can be used to account for any bit leaks in any reconciliation procedure. Many unstated and strictly invalid assumptions are used in QKD security proofs, as outlined in this paper and in [61], any of which would invalidate any claim to proven security. ​The security of QKD protocols requires a substantial amount of further careful study. ​
\\

Objection I:  It has not been explained how an imperfect key would affect QKD security when used for error correction​ and what the overall complexity security becomes when used in a conventional symmetric-key
cipher.

Answer:  The first question is a major open problem in QKD security theory. The second occurs when a QKD key is used in ciphers such as AES; however, it is not very relevant because an imperfect key can only weaken the complexity security compared to a uniform key. The substantive question is what the complexity security becomes if the imperfect seed key is changed more often compared to a uniform key. Both problems appear to be very difficult and seemingly not amenable to analysis.

This paper never claims to address, let alone solve, all security problems associated with QKD-generated keys. The paper provides some fundamental results on the security of any key generation scheme, quantum as well as classical, and notes some serious unsolved problems. Whoever claims security has the burden of providing a valid proof.

\section*{Acknowledgment}

I would like to thank Greg Kanter for his discussions that helped clarify some of the issues treated in this paper. My cryptography research has been supported by the Defense Advanced Research Project Agency and the United States Air Force.

\end{document}